\documentclass[useAMS,usenatbib]{mn2e}
\pdfoutput=1
\usepackage{xspace}
\usepackage{amssymb}
\usepackage{graphicx} 
\usepackage{amsmath}
\usepackage[T1]{fontenc}

\newcommand{\kms}{km~s$^{-1}$\xspace} 
\newcommand{\msun}{M$_{\odot}$\xspace} 
\newcommand{\dnu}{$\Delta \nu$\xspace} 
\newcommand{\numax}{$\nu_{\mathrm{max}}$\xspace} 
\newcommand{\teff}{$T_{\mathrm{eff}}$\xspace} 
\newcommand{\aFe}{[$\alpha$/M]\xspace} 
\newcommand{\al}{$\alpha$} 
\title{Young alpha-enriched giant stars in the solar neighbourhood}

\author[M. Martig et al.]{Marie Martig$^{1}$, Hans-Walter Rix$^{1}$, Victor Silva Aguirre$^{2}$, Saskia Hekker$^{3,2}$, Benoit Mosser$^4$,
\newauthor  Yvonne Elsworth$^5$, Jo Bovy$^6$, Dennis~Stello$^{7,2}$, Friedrich Anders$^8$, Rafael A. Garc\'{\i}a$^9$,
\newauthor   Jamie Tayar$^{10}$, Tha\'ise S. Rodrigues$^{11, 12, 13}$, Sarbani Basu$^{14}$, Ricardo Carrera$^{15, 16}$, 
\newauthor Tugdual Ceillier$^9$,  William J. Chaplin$^{5, 2}$, Cristina Chiappini$^8$, Peter M. Frinchaboy$^{17}$,
\newauthor  D. A.  Garc\'{\i}a-Hern\'{a}ndez$^{15, 16}$, Fred R. Hearty$^{18}$,  Jon Holtzman$^{19}$, Jennifer A. Johnson$^{10}$, 
\newauthor Steven R. Majewski$^{20}$, Savita Mathur$^{21}$, Szabolcs~M{\'e}sz{\'a}ros$^{22}$, Andrea Miglio$^{2,5}$, 
\newauthor David Nidever$^{23}$, Kaike Pan$^{24}$, Marc Pinsonneault$^{10}$, Ricardo P.  Schiavon$^{25}$,
\newauthor  Donald P. Schneider$^{18,26}$, Aldo Serenelli$^{27}$, Matthew Shetrone$^{28}$, Olga Zamora$^{15, 16}$\\
$^1$Max-Planck-Institut f\"{u}r Astronomie, K\"{o}nigstuhl 17, 69117 Heidelberg, Germany\\
$^2$Stellar Astrophysics Centre, Department of Physics and Astronomy, Aarhus University, Ny Munkegade 120, DK-8000 Aarhus C, Denmark\\
$^3$Max Planck Institut f\"ur Sonnensystemforschung, Justus-von-Liebig-Weg 3, 37077 G\"ottingen, Germany\\
$^4$LESIA, CNRS, Universit\'e Pierre et Marie Curie,  Universit\'e Denis Diderot, Observatoire de Paris, 92195 Meudon Cedex, France\\
$^5$School of Physics and Astronomy, University of Birmingham, Edgbaston, Birmingham B15 2TT, UK\\
$^6$Institute for Advanced Study, Einstein Drive, Princeton, NJ 08540, USA \\
$^7$Sydney Institute for Astronomy (SIfA), School of Physics, University of Sydney, NSW 2006, Australia\\
$^8$Leibniz-Institut für Astrophysik Potsdam (AIP), an der Sternwarte 16, 14482, Potsdam, Germany\\
$^9$Laboratoire AIM, CEA/DSM -- CNRS - Univ. Paris Diderot -- IRFU/SAp, Centre de Saclay, 91191 Gif-sur-Yvette Cedex, France\\
$^{10}$Ohio State University, Dept. of Astronomy, 140 W. 18th Ave., Columbus, OH 43210, USA\\
$^{11}$Osservatorio Astronomico di Padova -- INAF, Vicolo dell'Osservatorio 5, I-35122 Padova, Italy\\
$^{12}$Dipartimento di Fisica e Astronomia, Universit\`a di Padova, Vicolo dell'Osservatorio 2, I-35122 Padova, Italy\\
$^{13}$Laborat\'orio Interinstitucional de e-Astronomia -- LIneA, Rua Gal.\ Jos\'e Cristino 77, Rio de Janeiro, RJ -- 20921-400, Brazil\\
$^{14}$Department of Astronomy, Yale University, PO Box 208101,New Haven, CT 06520-8101, USA\\
$^{15}$Instituto de Astrof\'{\i}sica de Canarias, V\'{\i}a Lactea s/n, E-38200, La Laguna, Tenerife, Spain\\
$^{16}$Departamento de Astrof\'{\i}sica, Universidad de La Laguna, (ULL), E-38206 La Laguna, Tenerife, Spain\\
$^{17}$Department of Physics \& Astronomy, Texas Christian University, Fort Worth, TX, 76129, USA \\
$^{18}$Department of Astronomy and Astrophysics, The Pennsylvania State University, University Park, PA, 16802, USA\\
$^{19}$New Mexico State University, Las Cruces, NM 88003, USA\\
$^{20}$Department of Astronomy, University of Virginia, Charlottesville, VA 22904-4325, USA\\
$^{21}$Space Science Institute, 4750 Walnut street Suite\#205, Boulder, CO 80301, USA\\
$^{22}$ELTE Gothard Astrophysical Observatory, H-9704 Szombathely, Szent Imre herceg st. 112, Hungary\\
$^{23}$Department of Astronomy, University of Michigan, Ann Arbor, MI, 48104, USA\\
$^{24}$Apache Point Observatory, PO Box 59, Sunspot, NM 88349, USA\\
$^{25}$Astrophysics Research Institute, Liverpool John Moores University, IC2, Liverpool Science Park 146 Brownlow Hill Liverpool L3 5RF, UK\\
$^{26}$Institute for Gravitation and the Cosmos, The Pennsylvania State University, University Park, PA 16802, USA\\
$^{27}$Institute of Space Sciences (CSIC-IEEC) Campus UAB, Torre C5 parell 2 Bellaterra, 08193, Spain\\
$^{28}$University of Texas at Austin, McDonald Observatory 82 Mt. Locke Rd. McDonald Observatory, TX 79734 USA\\
}
\begin{document}
\maketitle
\vspace{-2cm}
\begin{abstract} 
We derive age constraints for 1639 red giants in the APOKASC sample for which seismic parameters from \textit{Kepler}, as well as effective temperatures, metallicities and [$\alpha$/Fe] values from APOGEE DR12 are available. We investigate the relation between age and chemical abundances for these stars, using a simple and robust approach to obtain ages. We first derive stellar masses using standard seismic scaling relations, then determine the maximum possible age for each star as function of its mass and metallicity, independently of its evolutionary stage.  While the overall trend between maximum age and chemical abundances is a declining fraction of young stars with increasing [$\alpha$/Fe], at least 14 out of 241 stars with [$\alpha$/Fe] $>0.13$ are younger than 6 Gyr. Five stars with [$\alpha$/Fe] $\geq 0.2$ have ages below 4 Gyr. We examine the effect of modifications in the standard seismic scaling relations, as well as the effect of very low helium fractions, but these changes are not enough to make these stars as old as usually expected for \al-rich stars (i.e., ages greater than 8--9 Gyr). Such unusual \al-rich young stars have also been detected by other surveys, but defy simple explanations in a galaxy evolution context.
\end{abstract}


\section{Introduction} 
Stellar ages and chemical abundances are amongst the key parameters that are used to constrain models of the formation and evolution of the Milky Way \cite[e.g.,][]{Schonrich2009, Minchev2013, Bird2013,Stinson2013}. In the absence of accurate age determinations for extended samples of stars, the abundance in \al-elements, [$\alpha$/Fe], may serve as a proxy for age. For instance, \cite{Bovy2012} decomposed the disc of the Milky Way into mono-abundance populations (in the  [$\alpha$/Fe] versus [Fe/H] plane), but any evolutionary interpretation must rely on a relation between abundances and age. Simulations by \cite{Stinson2013} showed that mono-abundance and mono-age populations are roughly equivalent. This equivalence still must be tested from an observational point of view, even though it has often been demonstrated that the large majority of \al-rich stars are indeed older than 8--9 Gyr (see for instance \citealp{Haywood2013,Bensby2014,Bergemann2014}).

A major difficulty arises from the fact that stellar ages must be indirectly inferred, contrary to stellar masses and radii, that can be measured for stars in binary systems, or using interferometry. Stellar ages are always model-dependent, and are often based upon the location of stars with respect to theoretical isochrones in the Hertzsprung--Russell (H--R) diagram (see \citealp{Soderblom2010} for a general review on stellar ages). Most of the early studies simply adopted the age of the isochrone closest to the data point \citep{Edvardsson1993, Ng1998, Feltzing2001}, while modern methods usually use some form of Bayesian parameter estimation, by computing the likelihood of stellar parameters versus a wide grid of theoretical isochrones \citep[e.g.,][]{Pont2004,Jorgensen2005,daSilva2006,Schonrich2014}. This technique is very powerful for stars near the main-sequence turn-off and on the sub-giant branch, where isochrones of different ages are clearly separated in the H--R diagram. On the red giant branch, isochrones are close to each other, rendering age determination difficult. Giant stars are, however, extremely important observational targets, as they cover a large range of ages and metallicities, and they are observable out to large distance. They constitute for instance the main targets for the APOGEE survey (\citealp{Zasowski2013}, Majewski et al., in preparation).

The prospects for age determination for red giants have been considerably enhanced with the advent of asteroseismic observations by the  CoRoT \citep{Baglin2006} and  \textit{Kepler} \citep{Borucki2010} space missions, that  probe the internal structure of stars and provide additional constraints on their properties. Solar-like oscillations have been detected in thousands of red giants, both by \textit{Kepler} and CoRoT  \citep[e.g.,][]{DeRidder2009, Hekker2009,  Bedding2010, Mosser2010,Hekker2011,Stello2013}, for stars out to 8 kpc from the Sun \citep{Miglio2013}. Solar-like oscillations are pulsations that are stochastically excited by convective turbulence in the stellar envelope \citep[e.g.,][]{Goldreich1977, Samadi2001}. 
These oscillation modes are regularly spaced in frequency, and can be described by  two global asteroseismic parameters, \dnu and \numax. Scaling relations connect these two seismic parameters to stellar mass, radius and effective temperature (see Section \ref{sec:kepler} for a description of these scaling relations). It is either possible to directly use scaling relations to determine stellar masses \citep{SilvaAguirre2011,Chaplin2011}, or to combine seismic information with theoretical isochrones to help lift some of the degeneracies we mentioned earlier  \citep[e.g.,][]{Stello2009,Kallinger2010,Basu2010, Quirion2010, Casagrande2014}. The latter technique is called "grid-based modelling"; one of its advantages is that it provides an estimate of stellar ages (ages cannot be directly derived from the scaling relations). Typical uncertainties on grid-based ages are below 30\% \citep[e.g.,][]{Gai2011,Chaplin2014}, but these ages are model-dependent, since they depend on the assumptions used to build the theoretical isochrones. 

The goal of this paper is to study the relation between age and chemical abundance for a sample of red giant stars that have been observed both by \textit{Kepler} (and thus have seismic parameters measured) and by APOGEE (\teff and chemical abundances are obtained from high-resolution near-infrared spectra). To derive ages, we use a simple technique that provides an upper limit on stellar ages, in a robust manner that minimizes as much as possible the effects of model-dependence that are characteristic of grid-based modelling. We first determine a minimum mass for each star using the seismic scaling relations,  then translate that minimum mass into a maximum age, which is insensitive to the evolutionary stage of each star. We also vary model assumptions to assess the robustness of our upper limits.

Our results confirm the expectations that most \al-poor stars have young ages, and that the fraction of young stars decreases with increasing [$\alpha$/Fe]. However, we also identify 14 stars that are both \al-rich and younger than 6 Gyr, which are not predicted by chemical evolution models of the Galaxy. 

We start by describing the APOKASC survey and the data we use in Section 2. In Section 3,  we justify the general motivation for our approach, and then explain in Sections 4 and 5 how we constrain masses and ages. The relation between age and chemical abundances is presented in Section 6, before discussing in Section 7 the robustness of our age and mass measurements. We conclude with a  brief discussion on the possible nature of the seemingly \al-rich young stars.

\section{The APOKASC sample}
APOKASC results from the spectroscopic follow-up by  APOGEE (Apache Point Observatory Galactic Evolution Experiment, Majewski et al., in preparation, as part of the third phase of the Sloan Digital Sky Survey, SDSS-III; \citealp{Eisenstein2011}) of stars with asteroseismology data from the \textit{Kepler} Asteroseismic Science Consortium (KASC).  The first APOKASC data release was presented in \cite{Pinsonneault2014}. The catalogue contains seismic and spectroscopic information for 1916 giants. In the original catalogue, the spectroscopic information corresponds to APOGEE's Data Release 10 (DR10; \citealp{Ahn2014}). For  this paper, we keep the same original sample of 1916 stars and their seismic parameters, but update their \teff and abundances to DR12 values \citep{Alam2015}. DR12 provides a number of improvements over DR10: the line list has been updated, the abundances of model atmospheres used to calculate synthesized spectra that are fitted to the observation \citep{Meszaros2012} are now consistent with the abundances used in the synthesis, and individual abundances for 15 elements are now computed.

\subsection{Seismic parameters from \textit{Kepler}}\label{sec:kepler}

Solar-like  oscillation modes are regularly spaced in frequency, and can be described by  two global asteroseismic parameters, \dnu and \numax, which can be used to measure stellar masses and radii. The large frequency separation, \dnu, is the frequency separation of two modes of same spherical degree and consecutive radial order. It is related to the sound travel time from the centre of the star to the surface, and depends on the stellar mean density \citep{Tassoul1980,Ulrich1986,Kjeldsen1995},
\begin{equation}\label{eq:dnu}
\Delta \nu \propto \rho^{1/2} \propto M^{1/2} R^{-3/2}\ .
\end{equation}
The power spectrum of the oscillations usually has a Gaussian-shaped envelope. The frequency of maximal oscillation power is called \numax, and is related to the acoustic cut-off frequency \citep{Brown1991}. In the adiabatic case,  and for an ideal gas, \numax mainly depends on surface gravity and temperature \citep{Kjeldsen1995,Belkacem2011},
\begin{equation} \label{eq:numax}
\nu_{\mathrm{max}} \propto g T_{\mathrm{eff}}^{-1/2} \propto  M R^{-2}T_{\mathrm{eff}}^{-1/2}\ .
\end{equation}

The $\sim 2000$ giants have been observed by \textit{Kepler} in long cadence mode, i.e., with a 30 minute interval \citep[e.g., ][]{Jenkins2010}. The light curves correspond to 34 months of data (Q0--Q12). They were prepared as described in \cite{Garcia2011}, and their power spectra were analysed with five different methods to measure \numax and \dnu \citep{Huber2009,Hekker2010, Kallinger2010, Mathur2010,Mosser2011}. 
The \numax and \dnu values provided in the catalogue are the ones obtained with the OCT method from \cite{Hekker2010}, while the other techniques are used for an outlier rejection process (stars with \numax values that differ significantly from one technique to another are removed from the sample) and to estimate uncertainties on the measured parameters.

\subsection{Spectroscopic parameters from APOGEE}
APOGEE uses a multi-fibre spectrograph attached to the 2.5 m SDSS telescope \citep{Gunn2006} to collect high-resolution ($R= 22500$) H-band stellar spectra. After being treated by the APOGEE data reduction pipeline \citep{Nidever2015}, these spectra are fed to the APOGEE Stellar Parameter and Chemical Abundances Pipeline (ASPCAP; \citealp{Meszaros2013}, Garc\'{\i}a P\'{e}rez et al., in preparation), that works in two steps. First, the spectra are compared to a grid of synthetic spectra \citep{Meszaros2012, Zamora2015} to determine the main stellar parameters. This grid has six dimensions: \teff, log $g$, metallicity [M/H],	as well as enhancement in \al-elements \aFe, in carbon [C/M] and in nitrogen [N/M]. The best-fitting spectrum is found by performing a $\chi^2$ optimization, and the corresponding stellar parameters are assigned to the observed star. In the first step of the processing, the \al-elements  are not considered individually, but are varied together with respect to the solar value. As a second step, individual abundances for 15 elements (including six \al-elements:  O, Mg, Si, S, Ca, and Ti) are obtained by fitting small regions of the spectra around specific lines of interest; this second stage is only performed in DR12, not in DR10.

Finally, the raw \teff and abundances are calibrated as described for DR10 by \cite{Meszaros2013} and for DR12 by \cite{Holtzman2015}. The ASPCAP temperatures are compared and calibrated to the photometric temperatures calculated from the 2MASS  J-K$_s$ colour \citep[as in][]{Gonzalez2009}. In DR12, the abundances are calibrated in two steps, based on observations of stars in 20 open and globular clusters. Under the assumption of homogeneity within clusters, some small trends of abundance with temperature were noted for some of the abundances (see \citealp{Holtzman2015} for the amplitudes), and an internal calibration has been applied to remove these. On top of this, an external calibration was applied to [M/H] because the derived values for metal-poor clusters are higher than those found by other studies (this external calibration is only necessary for [M/H]$<-1$ and does not affect the stars we use in this study). No external calibration was applied to any other abundances, largely because of the challenge of finding homogeneous measurements of individual element abundances covering a wide range of parameter space.

Throughout the paper, we always use the recalibrated values of the effective temperature and element abundances. 
We adopt the value of the uncertainty on [Fe/H] as our metallicity uncertainty, and compute the uncertainty on \aFe by adding in quadrature the uncertainties on [Fe/H] and [O/H].

\subsection{Sample selection}
We draw on the APOKASC--DR12 giant stars sample, removing stars with relative uncertainties on \dnu and \numax greater than 10\% and stars with uncertainties on \aFe greater than 0.08 dex. We eliminate stars for which any of the ASPCAP flags are set to WARNING or BAD (this signals potential problems with the determination of spectroscopic parameters). 
We also remove the metal poor stars ([M/H] $< -1$) for which the standard seismic scaling relations might be less accurate \citep{Epstein2014}. Finally, we  exclude fast rotating stars  (14 rapid and 12 additional anomalous rotators), that might be accreting mass from a companion, and for which the surface properties might not correspond to the evolutionary stage \citep{Tayar2015}.  Out of the 1916 stars with seismic and spectroscopic information, 1639 stars remain; these objects form the sample used in this paper. For these stars, the uncertainties on [M/H] range from $\sim  0.05$ at [M/H] $= -1$  to 0.03 at [M/H] $= 0.3$,  the \aFe uncertainty is  0.05 on average, and the \teff uncertainty is 91~K for all stars.

\section{A robust approach to age estimates}

\begin{figure}
\centering 
\includegraphics[width=0.5\textwidth]{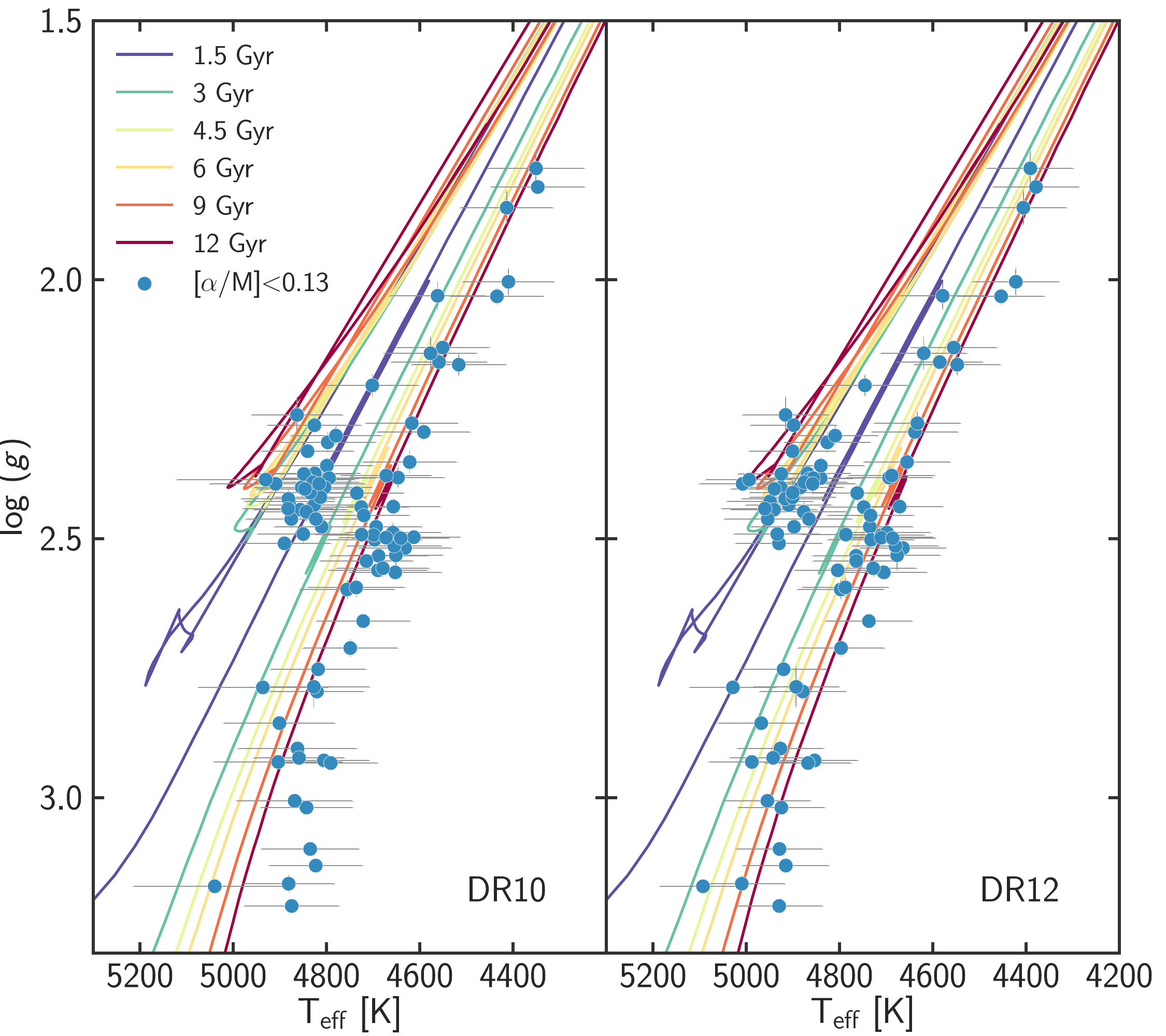}
\caption{Surface gravity as a function of effective temperature: comparison of APOKASC data (DR10 \teff and [M/H] on the left, DR12 on the right, log($g$) obtained from grid-based modelling using the seismic parameters) with a set of PARSEC isochrones. The data span a range of metallicity from -0.5 to -0.3. while the models are for a metallicity of -0.4 and are shown for ages from 1.5 to 12 Gyr. There is a systematic offset between data and theoretical isochrones for DR10 that is greatly reduced for DR12.}
\label{fig:HR}
\end{figure}

In addition to improved accuracy on element abundances, one of the reasons to update the spectroscopic parameters from DR10 to DR12  is an improvement in the \teff scale. Indeed, for DR10, there is a known offset between the observed \teff and standard isochrones, especially for $\mathrm{[M/H]} < -0.2$ \citep[see also][]{Meszaros2013,Pinsonneault2014,Bovy2014}. This offset is manifest in the left panel of  Figure \ref{fig:HR}, which presents a comparison of seismic log($g$) and DR10 \teff for a sample of APOKASC metal-poor stars ($-0.5 <$  [M/H] $< -0.3$) along a set of corresponding PARSEC v1.1\footnote{http://stev.oapd.inaf.it/cmd} isochrones \citep{Bressan2012} that have [M/H] $= -0.4$ and ages from 1.5 to 12 Gyr. We only show \al-poor stars to avoid potential issues with isochrones in the \al-rich regime ---  \aFe = 0.13 is a reasonable limit between \al-rich and \al-poor stars in APOKASC as can be seen in Figure \ref{fig:FeH_alpha}.  The observed temperatures are $\sim 100$~K  lower than the isochrones, with the magnitude of the offset increasing for more metal-poor stars. The right panel of Figure \ref{fig:HR} shows that this discrepancy is greatly reduced when DR12 \teff values are adopted.

Systematic offsets in effective temperature may arise both from the data and the models. For instance, it is known that temperatures derived from spectroscopy can be up to a few 100 K lower than temperatures computed from colours \citep{Johnson2002}. Temperatures derived using the infrared flux method for a sample of stars overlapping with the APOKASC sample  by \cite{Casagrande2014} as part of the SAGA survey (Str\"{o}mgren survey for Asteroseismology and Galactic Archaeology) are 90 $\pm 105$ K higher than the DR10 \teff.

The fact that DR12 \teff measurements are closer to the isochrones is reassuring, but the isochrone temperatures themselves could be affected by systematic uncertainties. Indeed,  the temperature on the red giant branch (RGB) in stellar models depends on many factors, including the treatment of convection, the atmospheric boundary conditions, the low-temperature opacities, and the equation of state \citep[e.g.,][]{Bressan2013}. In particular, convection is typically modelled within the framework of the mixing-length theory. In this theory, convection is described by a single number, the mixing length parameter, $\alpha_{\mathrm{MLT}}$, which is used to compute the typical distance an eddy can travel before losing its identity. In standard sets of isochrones, including the PARSEC ones, $\alpha_{\mathrm{MLT}}$ is fixed at a single value for all stellar models, and is calibrated such that the solar models reproduce observed properties of the Sun. However, the temperature on the RGB is highly sensitive to  $\alpha_{\mathrm{MLT}}$, and both 3D stellar atmosphere calculations  \citep{Trampedach2011,Magic2015} and calibrations on observations \citep{Bonaca2012} show that $\alpha_{\mathrm{MLT}}$ varies with temperature, gravity, and metallicity. Small changes in $\alpha_{\mathrm{MLT}}$ could easily shift the temperature of the RGB by a few \mbox{100 K}.
Finally, \teff is also affected by chemical abundances: on the RGB, \teff decreases  for a lower fraction of helium and for a higher fraction in \al-elements, in particular Mg and Si \citep{VandenBerg2012,VandenBerg2014}.
As a consequence, matching model isochrones of the RGB to data is highly non-trivial.

After exploring  these issues, we decided against using standard grid-based modelling techniques to compute ages, as we remained concerned that they may yield spurious ages because of the \teff offsets.
Instead, we adopt a simple approach that does not rely on precise comparisons with isochrones, which should be quite robust against \teff calibration issues. It simply consists in obtaining masses from the standard seismic scaling relations,  then translating these masses into ages using simple relations between stellar mass, metallicity and total lifetime, i.e., maximum age. Our aim is to provide strong and reliable constraints on the maximum age of each star. This approach also relies on the idea that masses and main-sequence lifetimes are more robustly understood than RGB temperatures. As demonstrated in \cite{Nataf2012},  a change of $\alpha_{\mathrm{MLT}}$ from 1.94 to 1.64, which would change \teff by several 100s K, only increases the stellar lifetime by 1\%.

\section{Mass estimates}

\subsection{Masses from standard seismic scaling relations}
The standard seismic scaling relations, Equations 1 and 2, can be combined to derive the mass of a star as:
\begin{equation} \label{eq:mass}
M= \left( \frac{\nu_{\mathrm{max}}}{\nu_{\mathrm{max,\odot}}}\right)^3\  \left( \frac{\Delta \nu}{\Delta \nu_{\odot}}\right)^{-4} \ \left( \frac{T_{\mathrm{eff}}}{T_{\mathrm{eff,\odot}}}\right)^{1.5} \ .
\end{equation}
We adopt  $T_{\mathrm{eff,\odot}}=5777$ K, $\nu_{\mathrm{max,\odot}}=3140\ \mu$Hz, $\Delta \nu_{\odot}=135.03\ \mu$Hz. The solar values  $\Delta \nu_{\odot}$ and $\nu_{\mathrm{max,\odot}}$ are the ones used to build the APOKASC catalogue and were obtained by \cite{Hekker2013} with the OCT method. 

We derive the mass uncertainty from the uncertainties on \numax, \dnu, and \teff, which have average values of 3.1\%, 2.4\%  and 1.9\%, respectively; this leads to an average mass uncertainty of 0.19 \msun (or 14\%). The uncertainties on \dnu and \numax are the main contributors to the error budget compared to uncertainties on \teff because of the higher exponent for these quantities in Equation \ref{eq:mass}. Temperature uncertainties are negligible here, except in the case of systematic offsets as discussed in the previous section. An increase of \teff by 100 K for all stars (see Figure \ref{fig:HR}) would increase the masses by up to 0.1 \msun.

\begin{figure}
\centering 
\includegraphics[width=0.5\textwidth]{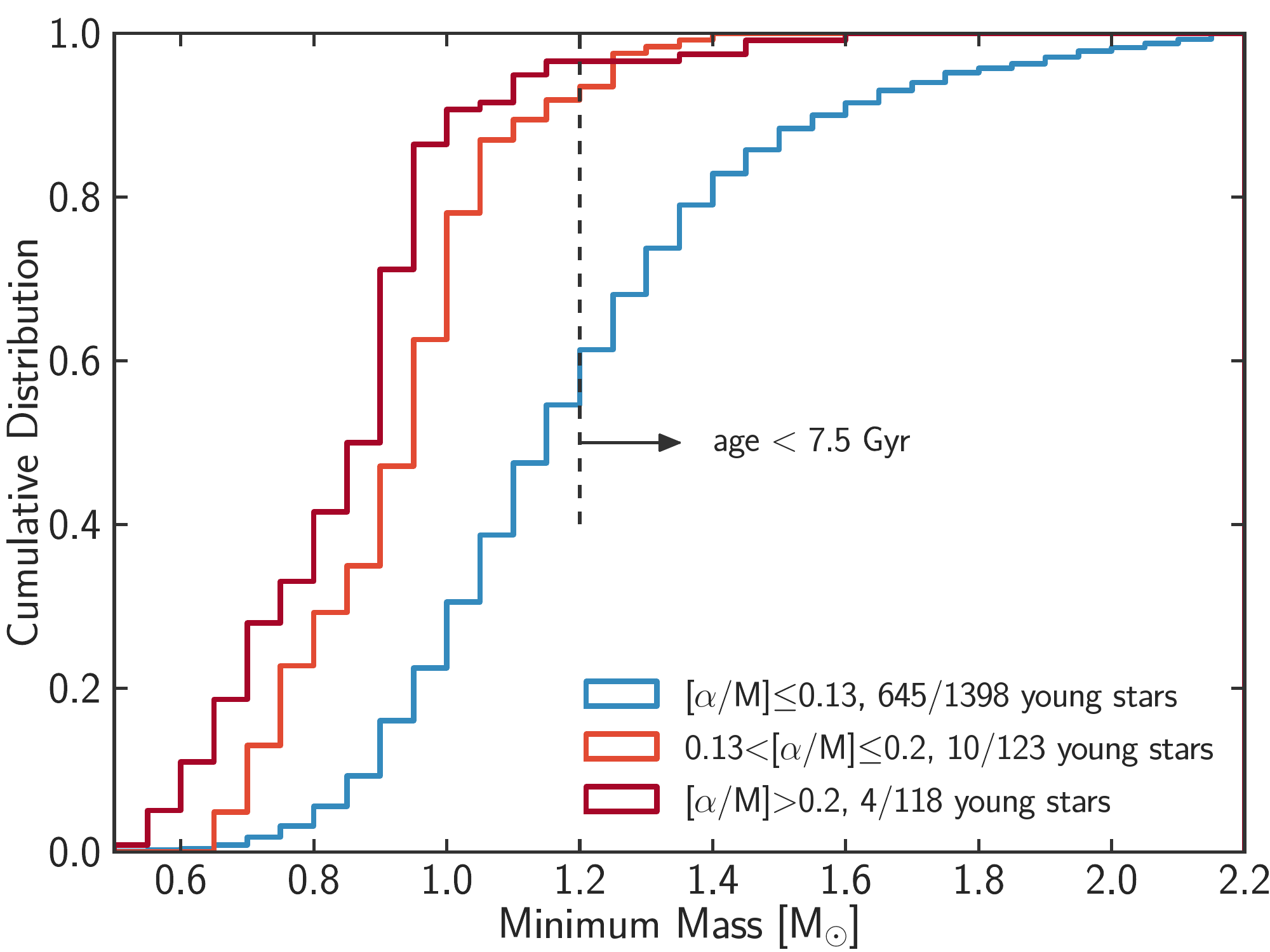}
\caption{Cumulative distribution of the "minimum mass" of stars in different bins of \aFe. We define the minimum mass as the lower bound of the 68\% confidence interval for each star. A minimum mass of 1.2 \msun translates into a maximum age of 7.5 Gyr, whatever the metallicity. Out of 241 stars with  \mbox{\aFe $>0.13$}, 14 are younger than 7.5 Gyr.}
\label{fig:mcumul}
\end{figure}

As we would like to provide an upper limit on stellar ages, we use the "minimum mass" of each star, defined as the 1$\sigma$ lower limit on the mass, which  translates into a "maximum age", defined as the corresponding 1$\sigma$ upper limit on the age. Figure \ref{fig:mcumul} displays the cumulative distribution of these minimum masses for stars in different bins of \aFe. As expected, the distribution is skewed to higher masses for stars with lower \aFe compared to \al-rich stars, indicating that \al-poor stars are on average younger than \al-old stars. There are, however, a few \al-rich stars (\aFe $>0.13$) with relatively high masses. As discussed in the next Section,  a minimum mass of 1.2 \msun corresponds to a maximum age of $\sim$ 7.5 Gyr. Figure \ref{fig:mcumul}  implies that amongst the 241 stars with  \aFe $>0.13$ there are 14 stars  younger than at least 7.5 Gyr. 

These seemingly remarkable stars constitute a small fraction of the overall sample and have mass uncertainties that are significant. Therefore, we must check whether their low estimates for the maximum age do not simply constitute the tail of the error distribution, with the true population of \al-rich stars being all old. To that aim, we select 241 mock giant stars from a set of PARSEC isochrones without mass loss on the RGB (in order to maximize stellar mass at a given age), with metallicities from $-0.9$ to 0.1, and ages uniformly distributed between 9 and 12.5 Gyr, which would be ages typically expected for \al-rich stars. We add errors to the masses by drawing randomly from the observed distribution of relative mass uncertainties ($\Delta$M/M), and compute the number of stars in that mock sample that would be wrongly identified as massive and young. We draw 1000 different such samples of 241 stars. We find that each sample contains on average $2 \pm 1.5$ stars with wrongly inferred minimum masses above 1.2 \msun. Only in 2\% of cases does the sample contain between 6 and 8 spurious young stars, and never more than 8. This result demonstrates that finding 14 young stars among 241 is inconsistent with a distribution of uniformly old stars given the observed mass uncertainties. Only a small fraction of these 14 stars could plausibly be erroneous inferences of massive stars. Section 7 describes some independent approaches to measure masses and ages that reinforce our trust in our mass estimates.

\subsection{Comparison with grid-based modelling results}

\begin{figure}
\centering 
\includegraphics[width=0.5\textwidth]{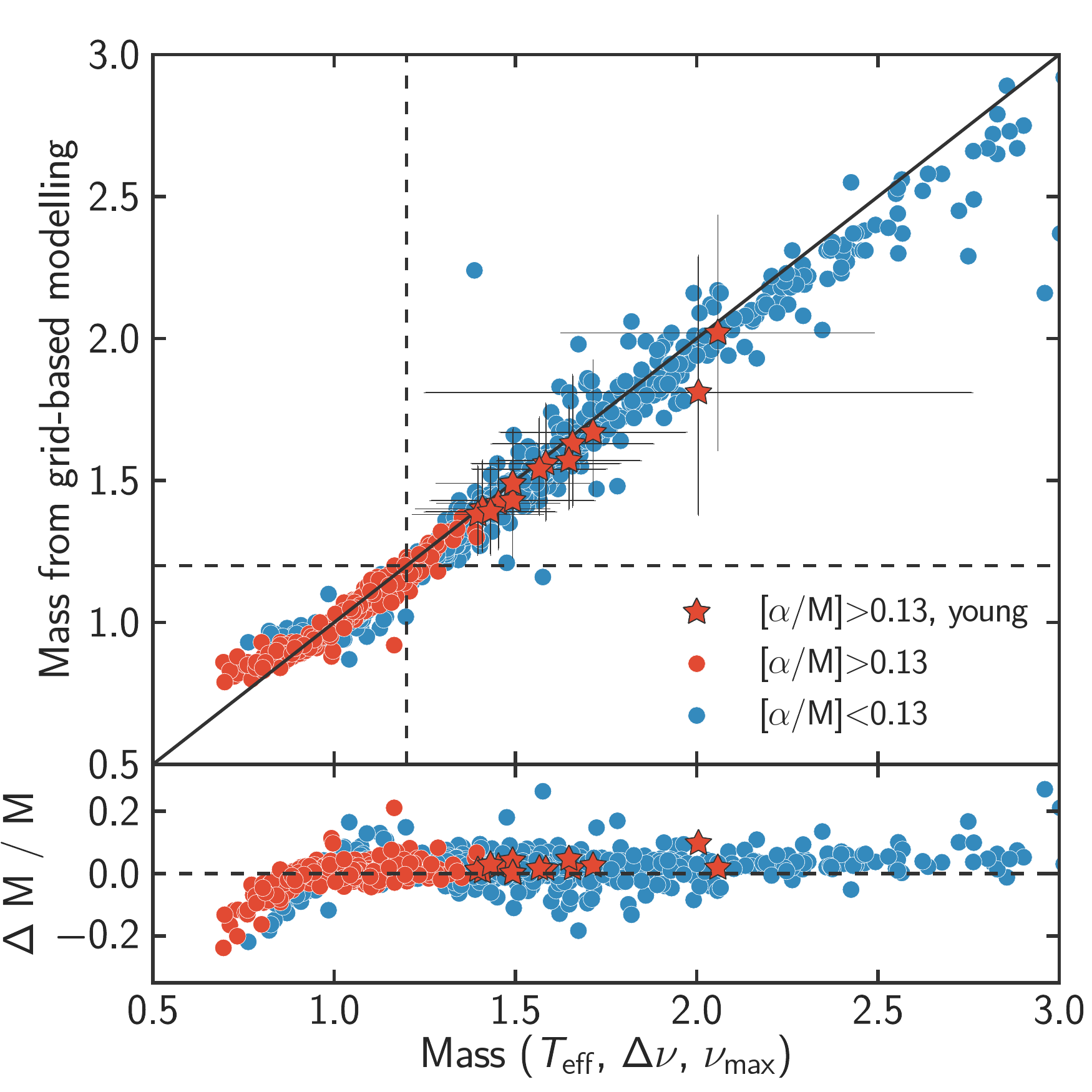}
\caption{Mass estimates obtained from grid-based modelling as a function of the mass directly derived from the seismic scaling relations (blue points: stars with \aFe$<$0.13, red points: stars with \aFe$>$0.13, red stars:  stars with \aFe$>$0.13 and a minimum mass greater than 1.2 \msun). Except at low masses, the two mass determinations agree well. The stars identified as massive and young (minimum mass $>1.2$ \msun) using scaling relations also have a high mass according to grid-based models.}
\label{fig:comp_grid}
\end{figure}

Using  the scaling relations directly is not the only way to derive masses for the APOKASC sample. The APOKASC catalogue contains values of masses and their associated \mbox{1-$\sigma$} uncertainties obtained from grid-based modelling. Grid-based modelling uses the seismic parameters combined with temperature and metallicity, and compares them to a grid of stellar models to derive the likelihood, or the posterior probability, of masses and ages. In the APOKASC catalogue, six different pipelines have been used to compute masses  \citep{daSilva2006,Stello2009,Basu2010, Kallinger2010, Serenelli2013,   Hekker2013}; their outputs have been combined as described in \cite{Pinsonneault2014}. These grid-based masses have been derived using DR10 \teff and [M/H], and not the DR12 versions that were not available yet when the analysis was performed.

Figure \ref{fig:comp_grid} presents a comparison of the mass derived from grid-based modelling as a function of the mass obtained using only the scaling relation. Except for low-mass stars, the two mass determinations agree well. The small systematic bias is due to the use of DR12 \teff in the scaling relations versus DR10 for the grid-based masses.  Compared to DR10, effective temperatures are on average higher by 50 K in DR12, and can be up to 150 K higher. This offset produces slightly increased masses for DR12-based determinations: the slight systematic bias seen in Figure \ref{fig:comp_grid}  disappears if  DR10 \teff values are used to derive masses from the scaling relations.

The red dots on this Figure represent the \al-rich stars; those with a minimum mass greater than 1.2 \msun according to scaling relations (the red stars) would also be classified that way from grid-based modelling. For these stars, the two mass determinations are in excellent agreement.

An often-cited advantage of grid-based modelling is  that the returned masses as a function of temperature, metallicity, and seismic parameters are consistent with stellar evolutionary models. By using only the scaling relation, all combinations of temperatures, radii, and masses are in principle possible. This means that the uncertainty on mass is reduced with grid-based modelling  \citep{Gai2011,Chaplin2014}, although \cite{Pinsonneault2014} show that the difference between techniques is less strong for red giants compared to main-sequence or sub-giants stars. We verify this conclusion  here: the average uncertainty on mass is 11.6\% with grid-based modelling, and 13.9\% using scaling relations.

Since we find no major systematic offset between the two mass determination techniques, for simplicity we choose to use the mass derived directly from the scaling relations, and its larger associated error bars. It also ensures that the masses and chemical abundances are all consistently derived from DR12 data.

\section{From masses to ages}

\subsection{Maximal age as a function of mass and metallicity}\label{sec:ages}
\begin{figure}
\centering 
\includegraphics[width=0.5\textwidth]{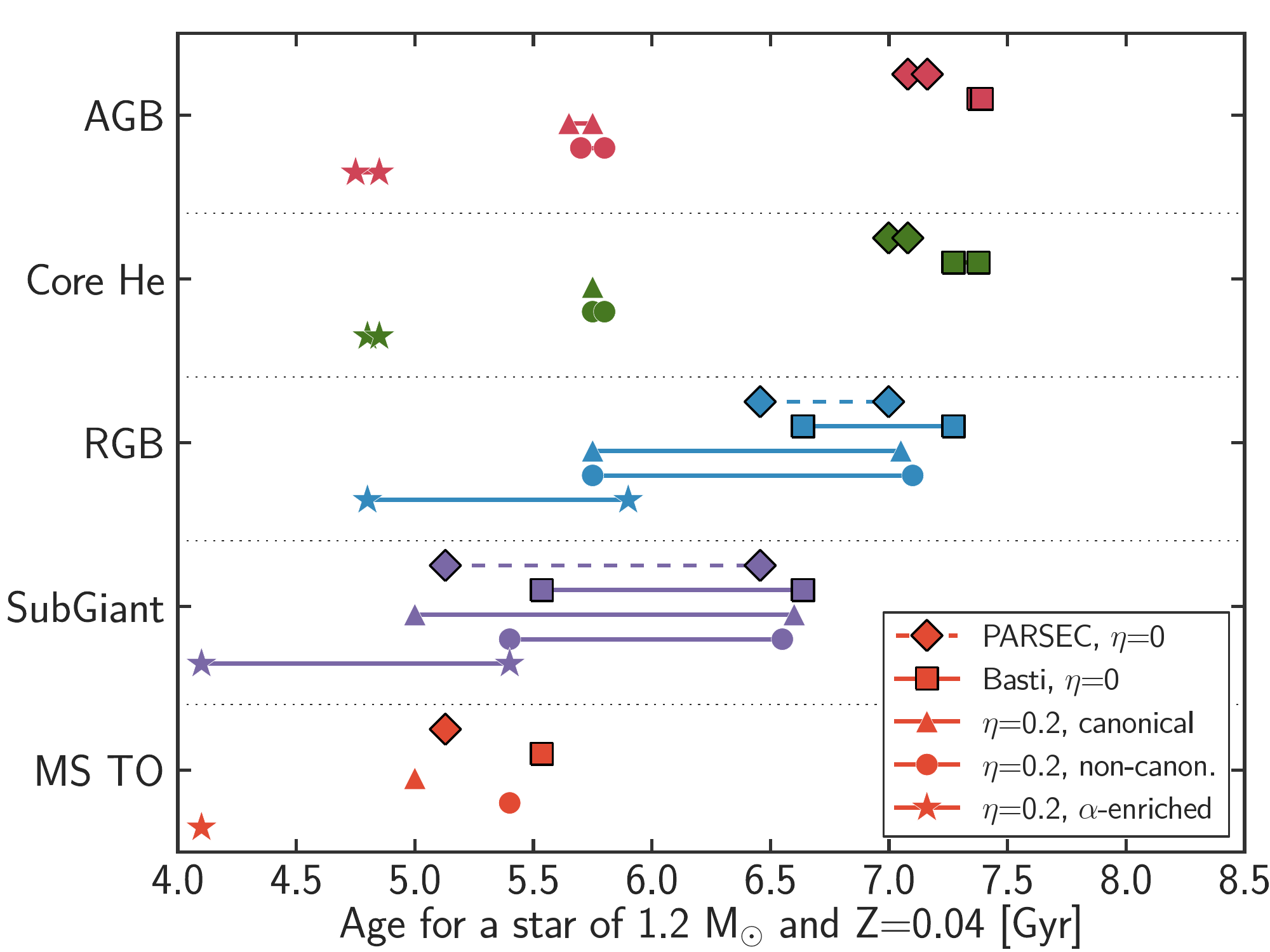}
\caption{Range of possible ages for a post main-sequence star with a mass of 1.2 \msun and $Z=0.04$. We show ages for the main-sequence turn-off, sub-giant branch, red giant branch, core helium burning phases, and asymptotic giant branch (from bottom to top --- points connected by lines represent the range of possible ages in each phase). For each of these phases, we show ages according to the PARSEC and BaSTI stellar evolution models. The diamonds and squares correspond to PARSEC and BaSTI models with no mass loss on the RGB (corresponding to $\eta=0$). The triangles, dots, and stars correspond to BaSTI with $\eta=0.2$ for canonical models, non-canonical models including core convective overshooting on the main sequence, and models with the same total metal content $Z$ but enriched in $\alpha$ elements. The highest possible ages (7.4 Gyr) are obtained on the AGB for models without mass loss, and differ by only a few 100 Myr between BaSTI and PARSEC models. Depending on the evolutionary stage of a star, its content in $\alpha$ elements and the physical model considered, the age of a post main-sequence star with perfectly measured mass could be up to $\sim$ 2 Gyr younger. }
\label{fig:stages}
\end{figure}

\begin{figure}
\centering 
\includegraphics[width=0.5\textwidth]{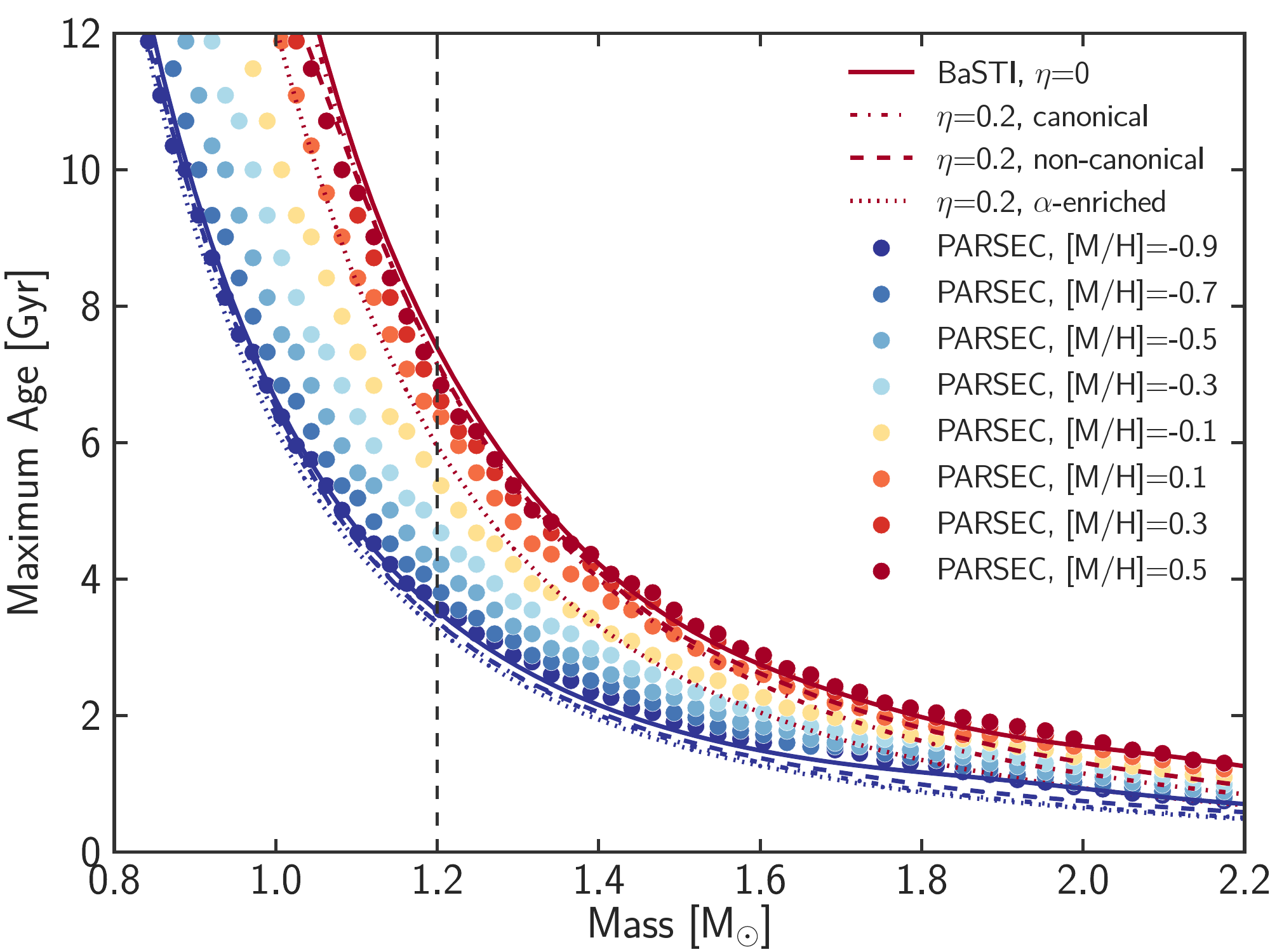}
\caption{Maximum age as a function of mass and metallicity for different stellar evolutionary models. The dots represent the values for the PARSEC isochrones for $\eta=0$ (no mass loss) and a range of metallicities, while the lines correspond to different sets of BaSTI isochrones for two extreme metallicities ($Z=0.002$ and $Z=0.04$) in red and blue (solid lines -- no mass loss,  dash-dotted lines -- $\eta=0.2$, canonical models,  dashed lines -- $\eta=0.2$, non-canonical models including core convective overshooting,  dotted lines -- $\eta=0.2$, models enriched in $\alpha$ elements). }
\label{fig:age_mass}
\end{figure}

Figure \ref{fig:stages} displays the range of possible ages for a post main-sequence  star of mass 1.2 \msun and metal content $Z=0.04$.  At a given mass, age increases with metallicity, so that any star more massive and/or more metal-poor than this will be younger. For our analysis, we use both PARSEC v1.1  and BaSTI \footnote{http://basti.oa-teramo.inaf.it/index.html} \citep{Pietrinferni2004,Pietrinferni2006} stellar evolutionary models. The adopted $Z$ corresponds to [M/H] $=0.4$ for BaSTI and 0.48 for PARSEC models because of different values of solar metallicities adopted by each model. This metallicity is representative for the most metal-rich stars in the APOKASC sample.  We present ages for different stages of evolution, from the main-sequence turn-off to the asymptotic giant branch for different sets of stellar evolutionary models. 

Two series of points in Figure \ref{fig:stages} correspond to models without any mass loss on the RGB (the diamonds for PARSEC, the squares for BaSTI). For both models, stars live for about 2 Gyr after leaving the main sequence, for a maximal age of 7--7.4 Gyr when they reach the AGB phase. BaSTI ages are systematically higher by a few 100 Myr, potentially because of a different helium fraction $Y$: for BaSTI, $Z=0.04$ corresponds to $Y=0.303$ while it corresponds to $Y=0.32$ for PARSEC, and an increased helium fraction reduces the stellar lifetime (see  Section \ref{sec:age_uncertainties} for a detailed discussion of the effect of the helium fraction on age).

This age of 7--7.4 Gyr corresponds to the maximal age for a 1.2 \msun star with $Z=0.04$. This also represents the maximal age for any star more massive and/or more metal poor than these values in our sample. 

Neglecting mass loss on the RGB provides an upper limit on ages, but is not the most realistic model for stellar evolution. The exact rates of mass loss on the RGB are uncertain, and depend on mass, luminosity, and temperature. They are usually parametrized with an efficiency parameter called $\eta$ \citep{Reimers1975}. Through a study of the mass difference between RGB and RC stars in two open clusters, \cite{Miglio2012} show that $\eta$ lies between 0.1 and 0.35; we adopt $\eta=0.2$. Figure  \ref{fig:stages} also shows ages for a 1.2 \msun star using the BaSTI isochrones with $\eta=0.2$ for the canonical model, for a non-canonical model that includes convective core overshooting on the main sequence and  semiconvective mixing in the core helium-burning phase, and for a canonical model with the same total $Z$ but enriched in \al-elements. These ages vary, because they do not correspond to the initially same star: a star with a mass of 1.2 \msun at the tip of the RGB actually started its evolution more massive, and is hence younger than a star with a similar mass at the base of the RGB. For the three models  with $\eta=0.2$  tested here, the highest possible ages for a 1.2 \msun star are found on the RGB (but not at the tip of the RGB), and are of about $\sim$ 7 Gyr, except for \al-enriched stars that could be 1 Gyr younger.

In the following, for a given mass and metallicity, we adopt as the maximum age the age on the AGB based on PARSEC models without mass loss. This is a less realistic model for stellar evolution compared to models including mass loss, but provides a more robust upper limit on ages (as seen in Figure \ref{fig:stages}). 
Figure \ref{fig:age_mass} demonstrates  how this maximum age depends on mass and metallicity. The series of dots correspond to PARSEC isochrones for a range of metallicities representative of the APOKASC sample, while the solid lines are the BaSTI maximal ages for two extreme metallicities ($Z=0.002$ and $Z=0.004$, roughly corresponding to the minimum and maximum values shown for PARSEC). As seen in Figure \ref{fig:stages}, the maximum age for a star of a given mass is similar between stellar evolution models, with differences up to a few 100 Myr at most. The dashed, dotted, and dash-dotted lines correspond to various BaSTI models including mass loss that yield smaller ages than the standard case without mass loss.

Because the age--mass relation becomes extremely steep for low masses, age determinations are difficult for these stars: a small error on mass or metallicity creates large variations in age. For this reason, we do not attempt to constrain ages for stars with $M<1.2$ \msun. 
For stars more massive than \mbox{1.2 \msun}, we determine a maximum age based on each star's minimum mass and metallicity by using the PARSEC isochrones with no mass loss (the dots in Figure \ref{fig:age_mass}). Instead of using the actual metallicity of each star, we use bins of 0.2 dex in metallicity as shown in Figure \ref{fig:age_mass}. In neighbouring bins, the change of age at a fixed mass is of the order of a few 100 Myr at most, thus potential small errors on the APOGEE metallicities have little influences on our results. The \al-enriched models produce  consistently younger ages, so that by not considering such models, we provide a conservative upper limit on ages.

\subsection{Age uncertainties due to the helium fraction}\label{sec:age_uncertainties}
\begin{figure}
\centering 
\includegraphics[width=0.5\textwidth]{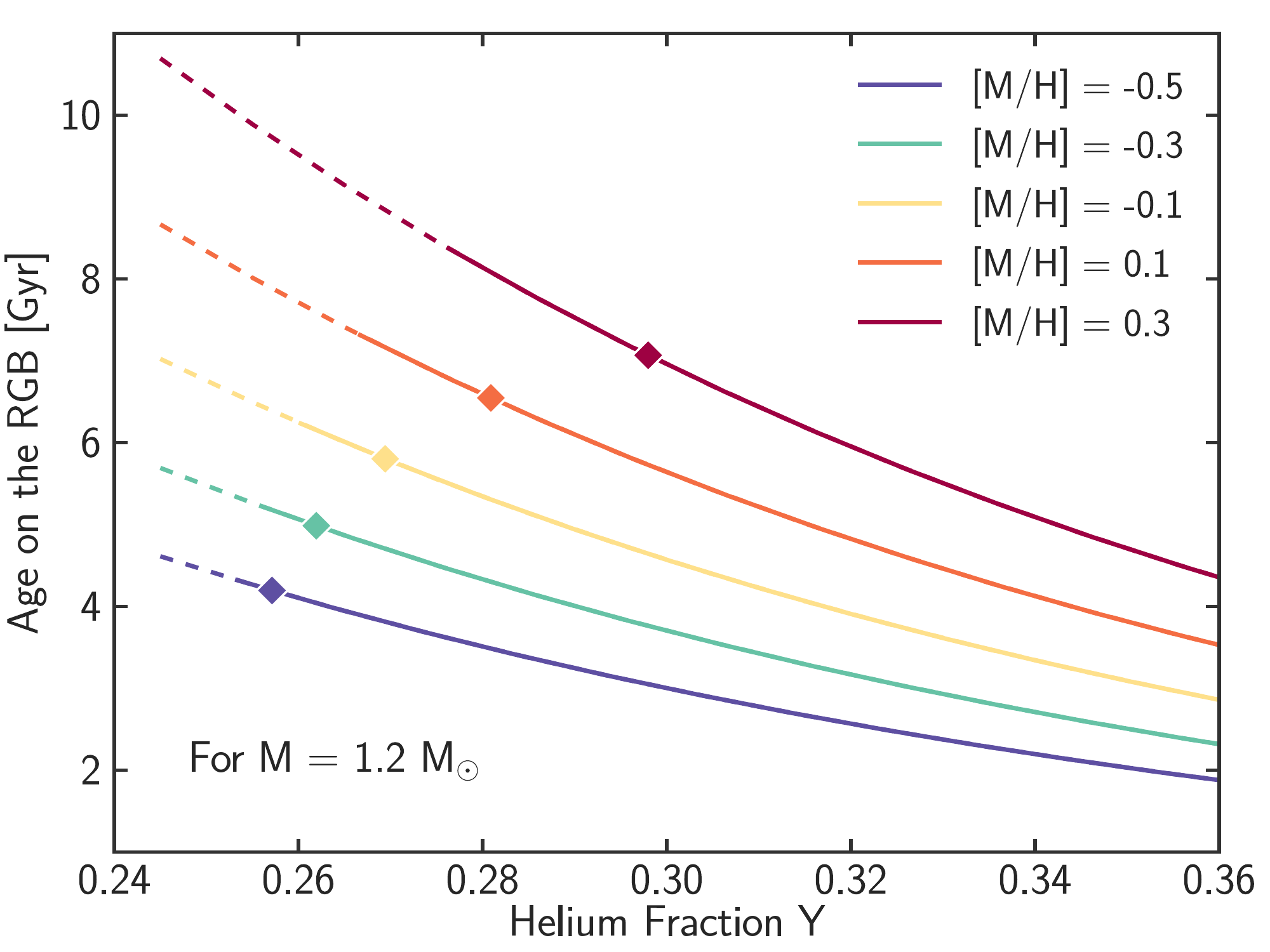}
\caption{Age on the RGB as a  function of the helium fraction (using Equation \ref{eq:helium}) for stars with a mass of 1.2 \msun and metallicities from -0.5 to 0.3 (representative of the range of metallicities for the young \al-rich stars in our sample). The dashed lines correspond to helium fractions for which $\mathrm{d}Y/\mathrm{d}Z <1$ (these are values of $Y$ lower than what is usually found), while the  diamonds represent the values adopted for the PARSEC isochrones ($\mathrm{d}Y/\mathrm{d}Z=1.78$)}
\label{fig:helium}
\end{figure}
Despite our efforts to arrive at an upper age limit that is robust against model uncertainties, there remains an additional uncertainty due to the unknown helium content of the stars. Direct measurements of the abundance in helium are limited to stars with \teff greater than 8000 K for which strong helium lines are present \citep{Valcarce2013}. The mass fraction in helium, $Y$, is, however, a critical parameter for stellar models because it has a strong influence on stellar ages. For a fixed stellar mass and metallicity, stars enriched in helium have a shorter lifetime \citep[e.g., ][]{Karakas2014}. Stellar evolution models usually assume that $Y$ varies as a function of $Z$ following a simple linear relation: $Y=Y_{p}+\frac{\mathrm{d}Y}{\mathrm{d}Z} Z$, where $Y_{p}$ is the primordial helium fraction ($Y_{p}=0.2485$ is adopted by \citealp{Bressan2012}), and $\mathrm{d}Y/\mathrm{d}Z$ is calibrated to reproduce the Sun's helium abundance. From combinations of observations and theory,  $\mathrm{d}Y/\mathrm{d}Z$ is usually found to be between 1 and 2.5 \citep{Ribas2000,Casagrande2007, Brogaard2012}, while values up to 10 are sometimes proposed \citep[e.g.,][]{Portinari2010}. In any case, it is  unclear if $Y$=f($Z$) should be a simple linear relation, or could have more complex behaviours. For instance, \cite{NatafGould2012} argue for an increased Y for \al-rich stars in the bulge.

To study the effect of a varying helium fraction on our stellar ages, we use the simple fitting formula provided by \cite{Nataf2012} to express stellar mass as a function of metallicity, initial helium abundance, and age upon reaching the RGB:
\begin{multline} \label{eq:helium}
\mathrm{log} \left( \frac{M}{M_{\odot}}\right) = 0.026 + 0.126 [\mathrm{M}/\mathrm{H}] -0.276\  \mathrm{log} \left( \frac{t}{10 \mathrm{Gyr}}\right) \\ -0.937 (Y-0.27)\ .
\end{multline}

Figure \ref{fig:helium} shows the age upon reaching the RGB as a function of $Y$ for stars with a mass of 1.2 \msun and metallicities from -0.5 to 0.3, representative of the range of [M/H] for the young \al-rich stars in our sample. This figure highlights the strong impact the $Y$ could have on age. However, for stars of 1.2 \msun, ages greater than 8 Gyr are only possible for stars with super-solar metallicities and unusually low helium fractions. These stars would correspond to  $\mathrm{d}Y/\mathrm{d}Z<1$, disfavoured by most studies. For stars with sub-solar metallicities, even a helium fraction as low as the primordial value still produces ages lower than 7 Gyr. Finally, for stars with a mass of 1.4 \msun (not shown in the Figure) and a sub-solar metallicity, ages are always lower than 4 Gyr, even with a primordial helium fraction.
\begin{table*}
\begin{center}
\caption{Properties of the 14 stars with \aFe>0.13 and young ages: all of these stars are younger than 6 Gyr. M$_{\mathrm{SR}}$ and M$_{\mathrm{GB}}$ are the masses as obtained directly from the scaling relations and from grid-based modelling. The last column corresponds to the maximum ages, expressed in Gyr. The uncertainty on \teff is 91 K in all cases. }\label{tab:ages}
\begin{tabular}{lccccccccc}
\hline
\hline
KIC ID &2MASS ID&\aFe&[M/H]&\teff[K]&\numax[$\mu$Hz]&\dnu[$\mu$Hz]&M$_{\mathrm{SR}}$[M$_{\odot}$]&M$_{\mathrm{GB}}$[M$_{\odot}$]&Age\\
\hline
9821622&2M19083615+4641212&0.26$\pm$0.05&-0.29$\pm$0.04&4780&63.72$\pm$1.49&5.91$\pm$0.19&1.71$\pm$0.26&1.67$^{+0.25}_{-0.22}$&$<$2.6\\
4143460&2M19101154+3914584&0.22$\pm$0.05&-0.24$\pm$0.04&4800&39.65$\pm$1.22&4.23$\pm$0.09&1.58$\pm$0.20&1.56$^{+0.21}_{-0.20}$&$<$3.1\\
4350501&2M19081716+3924583&0.21$\pm$0.05&-0.10$\pm$0.03&4824&143.83$\pm$3.69&11.03$\pm$0.24&1.65$\pm$0.20&1.57$^{+0.19}_{-0.17}$&$<$3.0\\
11394905&2M19093999+4913392&0.20$\pm$0.05&-0.44$\pm$0.04&4835&39.08$\pm$0.93&4.33$\pm$0.11&1.40$\pm$0.18&1.38$^{+0.17}_{-0.14}$&$<$4.0\\
9269081&2M19032243+4547495&0.20$\pm$0.05&-0.11$\pm$0.03&4807&25.30$\pm$1.27&2.83$\pm$0.10&2.06$\pm$0.43&2.02$^{+0.41}_{-0.41}$&$<$2.1\\
11823838&2M19455292+5002304&0.19$\pm$0.05&-0.40$\pm$0.04&4893&42.11$\pm$1.14&4.47$\pm$0.09&1.57$\pm$0.18&1.54$^{+0.18}_{-0.16}$&$<$3.1\\
5512910&2M18553092+4042447&0.19$\pm$0.05&-0.33$\pm$0.04&4898&39.98$\pm$1.33&4.24$\pm$0.09&1.66$\pm$0.22&1.63$^{+0.24}_{-0.22}$&$<$2.7\\
10525475&2M19102133+4743193&0.19$\pm$0.05&-0.18$\pm$0.03&4768&39.19$\pm$1.15&4.29$\pm$0.09&1.43$\pm$0.18&1.39$^{+0.16}_{-0.15}$&$<$4.7\\
9002884&2M18540578+4520474&0.16$\pm$0.04&-0.32$\pm$0.03&4187&4.82$\pm$0.17&0.78$\pm$0.07&2.00$\pm$0.75&1.81$^{+0.48}_{-0.43}$&$<$4.2\\
9761625&2M19093801+4635253&0.16$\pm$0.04&-0.17$\pm$0.03&4425&9.27$\pm$0.23&1.40$\pm$0.04&1.49$\pm$0.21&1.49$^{+0.19}_{-0.16}$&$<$4.3\\
11445818&2M19052620+4921373&0.16$\pm$0.04&-0.06$\pm$0.03&4767&37.05$\pm$1.37&4.07$\pm$0.10&1.49$\pm$0.23&1.43$^{+0.22}_{-0.20}$&$<$4.5\\
3455760&2M19374569+3835356&0.15$\pm$0.04&0.01$\pm$0.03&4609&47.61$\pm$1.04&4.85$\pm$0.10&1.49$\pm$0.16&1.49$^{+0.16}_{-0.14}$&$<$4.4\\
8547669&2M19052572+4437508&0.14$\pm$0.04&0.10$\pm$0.03&4492&27.40$\pm$0.70&3.22$\pm$0.08&1.41$\pm$0.18&1.40$^{+0.17}_{-0.15}$&$<$5.9\\
3833399&2M19024305+3854594&0.13$\pm$0.04&0.11$\pm$0.03&4679&37.80$\pm$0.87&4.13$\pm$0.09&1.45$\pm$0.17&1.42$^{+0.16}_{-0.16}$&$<$5.0\\
\hline
\end{tabular}
\end{center}
\end{table*}
While helium is in general a major uncertainty to take into account when computing ages, our upper limits do not strongly depend on it. Amongst the 14 young \al-rich stars that we find, only the most-metal rich ones (2 stars with [M/H]$\sim 0.1$) could be older than 8 Gyr if they have an extremely low $Y$. Such low $Y$ are extremely unlikely, especially for \al-rich stars \cite[see][]{NatafGould2012}.
 
\section{Ages and chemical abundances}
\begin{figure*}
\centering 
\includegraphics[width=\textwidth]{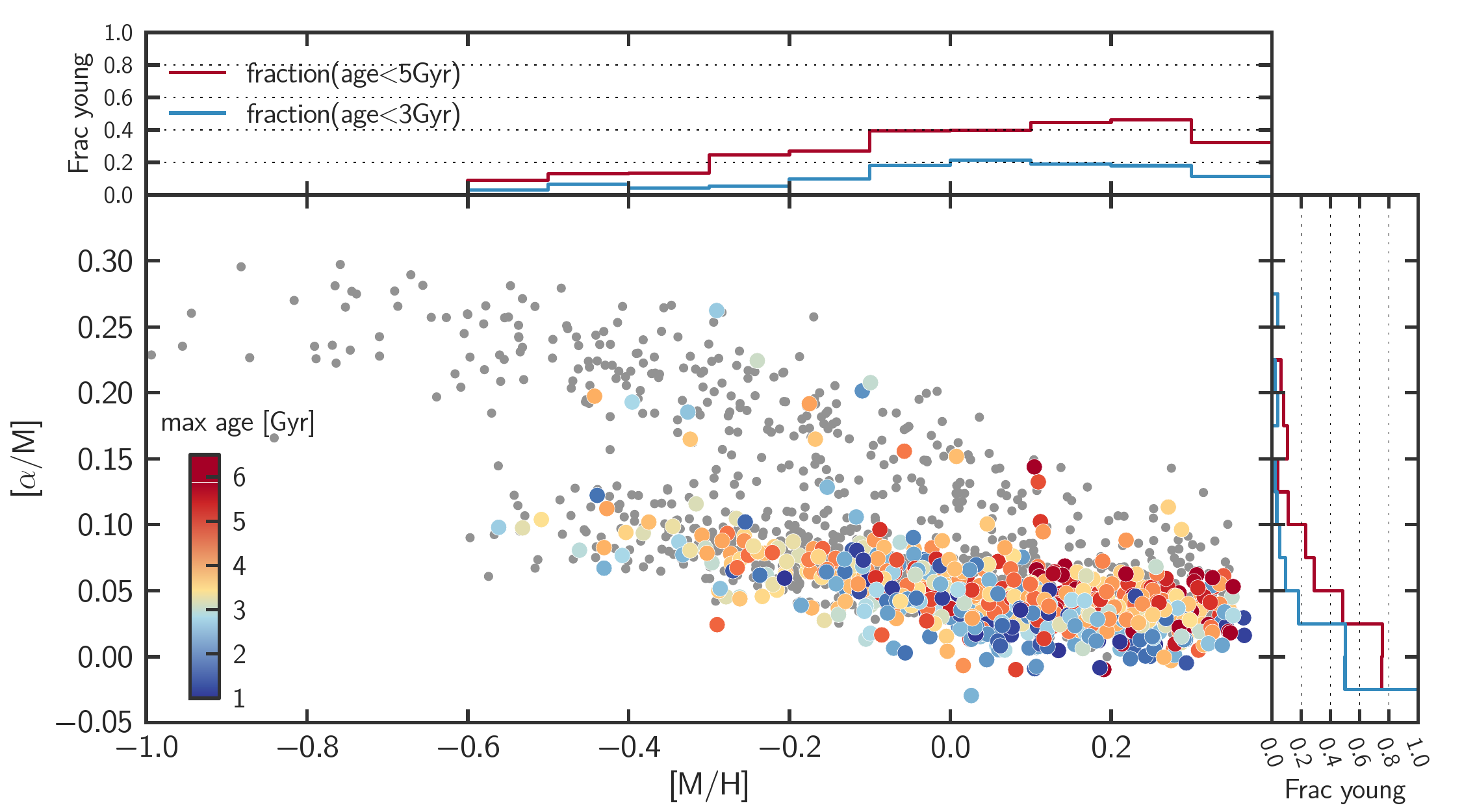}
\caption{Age and abundances for stars in our APOKASC sample. The main panel shows the distribution of the APOKASC sample in the \aFe versus [M/H] plane. Small grey dots represent stars for which ages are not measured (i.e., stars with a mass smaller than \mbox{1.2 \msun}), and coloured dots represent stars younger than 7 Gyr (the colour encodes the maximal age of each star). The two histograms show the fraction of stars younger than 5 and 3 Gyr in different bins of [M/H] (top) and \aFe (right)}
\label{fig:FeH_alpha}
\end{figure*}

Using the method described in Section \ref{sec:ages}, we compute a maximum age for all stars with a minimum mass above \mbox{1.2 \msun}, for which the age-mass relation is not too steep. Within our framework, this maximum age becomes a simple function of each star's minimum mass and [M/H].
We illustrate how this maximum age depends on abundances across the entire APOKASC sample in Figure \ref{fig:FeH_alpha}. This Figure shows the maximum age (colour code of the points)  as a function of \aFe and [M/H]. The grey dots are the stars with masses lower than 1.2 \msun, hence without an age determination; not all of these stars must be old since those with a low metallicity could have maximum ages of 4--6 Gyr.
The histograms on this Figure display the fraction of stars younger than 5 or 3 Gyr as a function of \aFe and [M/H]. 

As expected, we find that the highest fraction of young stars is found for  low \aFe. For \aFe$<0.025$, at least $\sim80$\% of stars are younger than 5 Gyr, and 50\% are younger than 3 Gyr. The fraction of young stars quickly decreases with increasing \aFe. By contrast, the relation between [M/H] and age is less strong. The [M/H] distribution of young stars is nearly flat for [M/H]$>-0.1$, and drops at low metallicity.

The most interesting outliers from these simple trends are the  14 stars that are quite \al-enhanced (with \aFe$>0.13$), yet have maximal ages smaller than 6 Gyr. They span the entire range of abundances in the \al-rich cloud, except the low metallicity tail ([M/H] $<-0.5$). Table \ref{tab:ages} lists the masses and maximum ages for these stars. In this list the three most \al-rich stars (with \aFe $>0.2$) have ages below $\sim$ 3 Gyr. If these age limits are correct, and if  \aFe represent the abundances of the material from which the stars formed, this result would be in strong contrast to the standard view in which \al-rich stars have ages greater than 8--9 Gyr.

\begin{figure*}
\centering 
\includegraphics[width=\textwidth]{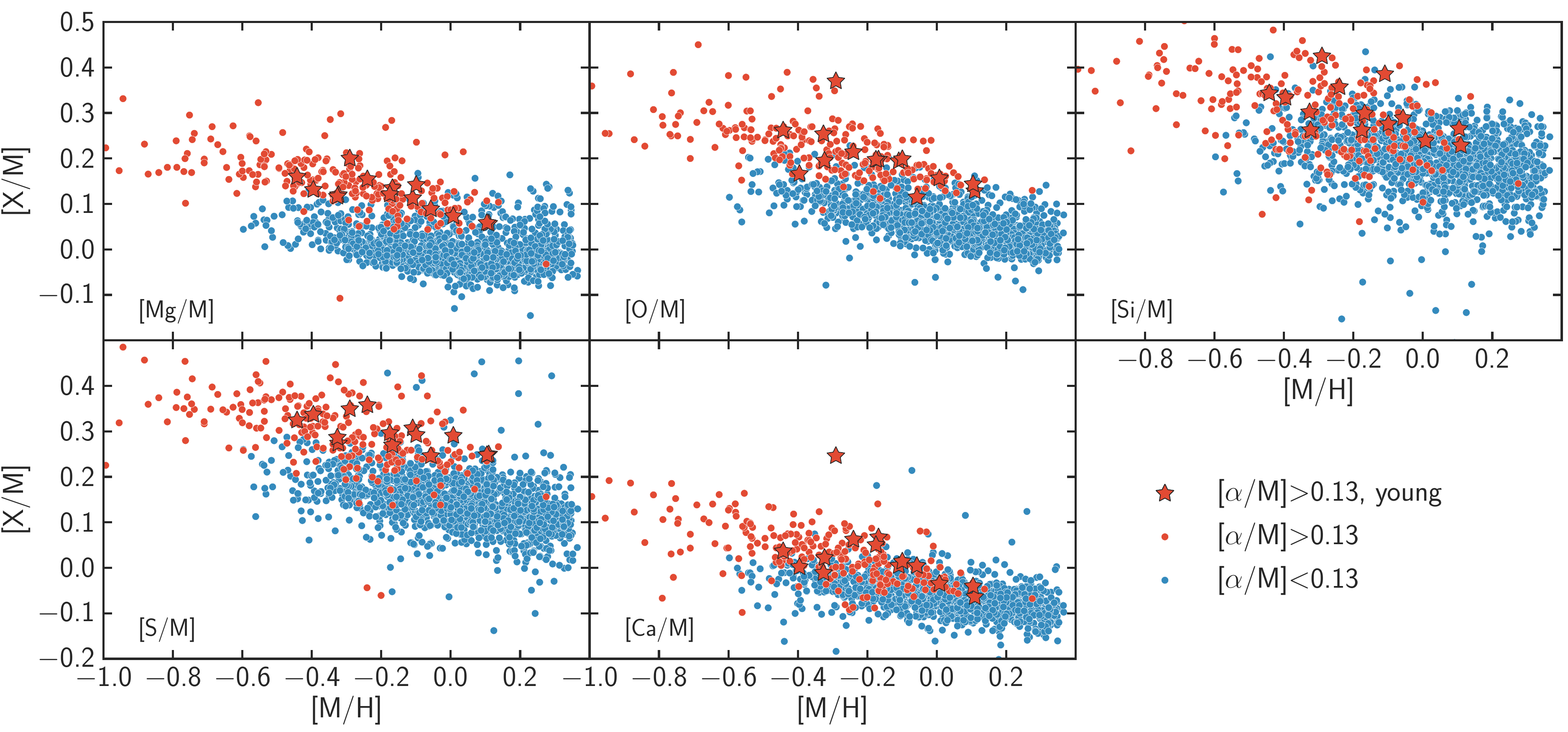}
\caption{Abundances in five \al-elements as a function of [M/H] for stars in the APOKASC sample. From top left to bottom right, the panels present [Mg/M], [O/M], [Si/M], [S/M], and [Ca/M] for \al-poor stars (blue dots), \al-rich stars (red dots) and \al-rich young stars (red stars). The young \al-rich stars generally follow the same abundance distribution as normal \al-rich stars. This demonstrates that our results are not driven by abnormalities in measurements of [\al/M]}
\label{fig:all_alphas}
\end{figure*}

One possible source of error in our analysis would be incorrect values of \aFe. However, visual inspection of the spectra and their ASPCAP fits reveals a good match. In addition, DR12 provides both a global value of \aFe (from the 6-parameter fit to the spectra), and abundances for individual \al-elements. Figure \ref{fig:all_alphas} presents the ratios [Mg/M], [O/M], [Si/M], [S/M], and [Ca/M]  as a function of [M/H] for the \al-poor stars, the \al-rich stars, and the \al-rich young stars\footnote{As discussed in \cite{Holtzman2015}, there is a zero-point issue for some of APOGEE abundances: the abundance in Si is too high compared to reference values, while the abundance in Ca is too low. This does not impact the present study, that just relies on the relative difference between \al-rich and \al-poor populations}. Not all elements trace faithfully the global \aFe value. For instance, there is a significant overlap between the values of [Si/M] for \al-rich and \al-poor stars. However, the \al-rich young stars behave similarly to the rest of the \al-rich stars. Only the most \al-rich of our 14 stars, KIC 9821622, is highly enriched in O and Ca, but the rest of the stars follow normal trends. The coherence between the global \aFe value and the individual abundances of \al-elements increases our confidence in the high \aFe values for our 14 young stars. For these stars the values of [M/H] from the global fit and of [Fe/H] from fits to individual Fe lines follow the same trend as for the whole sample, so that there is also no error on the metallicity determination.

\section{Robustness of our mass and age determinations}
Given that the existence of stars with ages lower than 3 Gyr and \aFe $>0.2$ would not only be interesting, but also surprising, we shall examine possible loopholes in our line of reasoning.
The seismic scaling relations are at the core of our study, as they are the basis of mass estimates, either directly or indirectly via grid-based modelling.
However, the scaling relations need to be critically examined. They are widely used but based on simplified assumptions about stellar structure. The radii derived from asteroseismology agree within 5\% with radii measured from interferometry \citep{Huber2012}, or using Hipparcos parallaxes \citep{SilvaAguirre2012}. Stellar masses are much more difficult to calibrate, but several studies suggest that seismic masses could be too high by 0.1--0.2 \msun, both at  low metallicity in the Milky Way halo \citep{Epstein2014} and in the open cluster NGC 6791, which has a super-solar metallicity \citep{Brogaard2012,Miglio2012}. Scaling relations could also slightly differ for Red Clump vs RGB stars \citep{Miglio2012}

\begin{figure}
\centering 
\includegraphics[width=0.5 \textwidth]{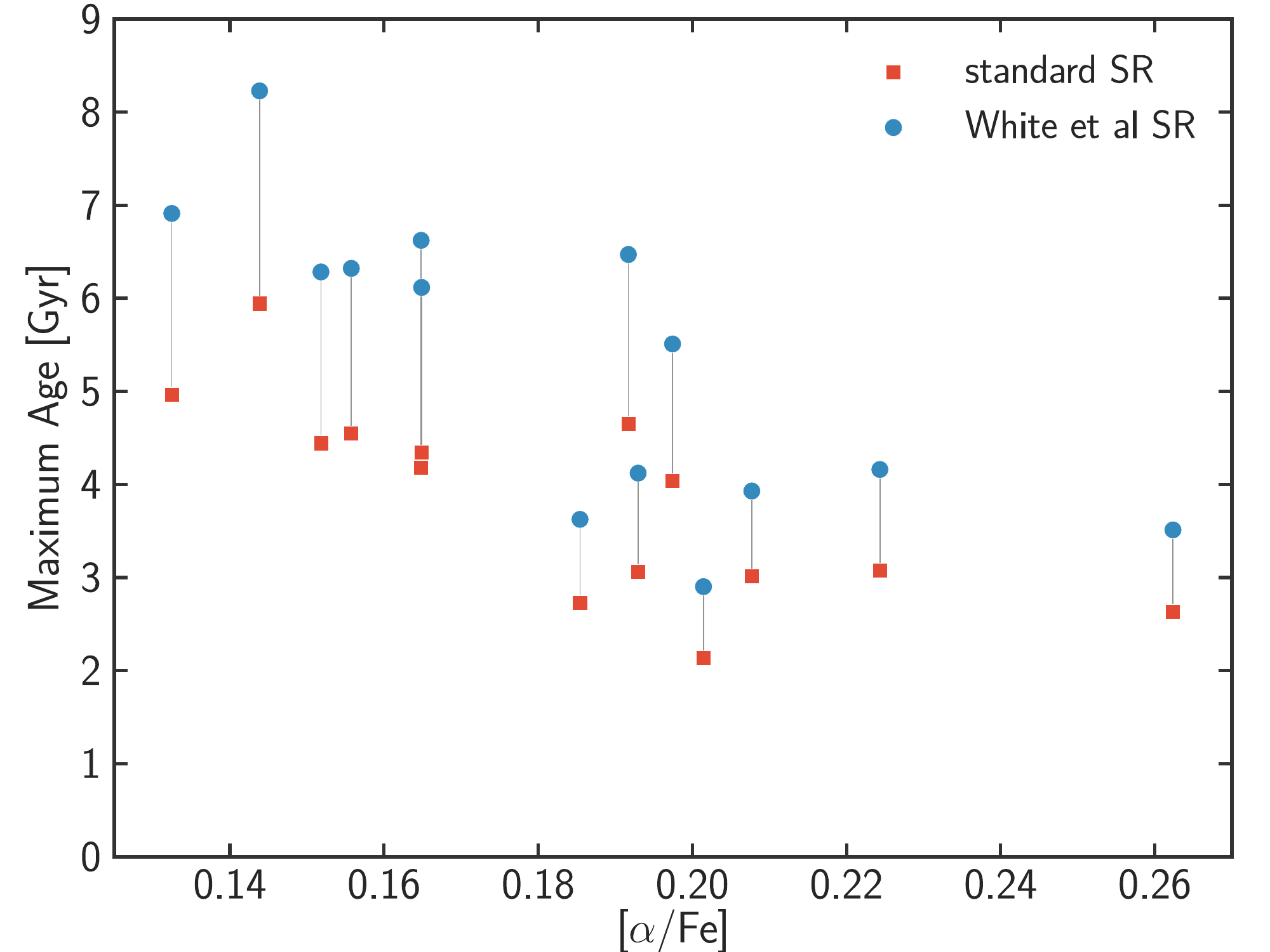}
\caption{Variations of the maximum age for our 14 young \al-rich stars, when deriving the mass from the standard scaling relations (red squares) or from the scaling relations modified by White et al. (2011) -- blue dots. Only for one star the maximum age becomes 8.2 Gyr, the others remain with maximum ages below 7 Gyr.}
\label{fig:white}
\end{figure}

\cite{White2011} propose a modification to the scaling relation between \dnu and the mean stellar density. They use theoretical evolutionary tracks to demonstrate that the relation between \dnu and $\rho$ depends on  effective temperature, mass, and metallicity, although the effect of mass and metallicity is less important. They propose a new scaling relation that is, however, only valid for \teff between 4700 and 6700 K. Since our stars are at the limit of this \teff range, we adopt a simple prescription, based on their Figures 4 to 6: \mbox{\dnu /\dnu$_{\odot} = 0.98 \sqrt{\rho/\rho_{\odot}}$}. This reduces all our masses by 8\%, which in turn has an effect of increasing ages, as shown in Figure \ref{fig:white}. The modifications in the masses are not enough to make all stars old; only one star has its age upper limit pushed to 8.2 Gyr. Six stars still have ages below 4.5 Gyr, including the four most \al-rich stars, while the other seven stars have maximum ages between 5.5 and 7 Gyr.

Another correction to the scaling relations was proposed by \cite{Mosser2013}, but the masses we compute using their Equation 29 are nearly identical to the masses derived from the standard scaling relations. The difference between both masses is always below 0.015 \msun. Finally, the visual inspection of the \textit{Kepler} light curves for these 14 stars did not reveal any abnormalities, and updated values of \numax and \dnu from longer time series (not yet available for the whole APOKASC sample) are consistent with the ones we used throughout the paper.

An additional constraint on mass can be obtained in some cases from the period spacing of mixed modes ($\Delta \Pi_1$). In red giants, the coupling between acoustic modes from the stellar envelope and gravity modes from the core produces mixed modes \citep{Beck2011}, and their period spacing can be used to distinguish between stars burning hydrogen in a shell  and stars burning helium in their core \citep{Bedding2011}. \cite{Mosser2014} show that the combination of $\Delta \Pi_1$ and \dnu can allow relatively precise diagnostics of a star's evolutionary stage. Table \ref{tab:type} presents values of $\Delta \Pi_1$ measured as in \cite{Mosser2012}, and the corresponding evolutionary stage, as in \cite{Mosser2014}. Not all stars have a measurable period spacing: for four stars the ambiguity remains. Three stars are on the RGB, seven are in the Red Clump, and one appears to be in the subflash phase, not yet in the clump. The high fraction of clump stars in our sample is expected given that clump stars are generally 3--4 times more numerous than stars on the upper RGB \citep{Nidever2014}.

To further confirm the young nature of some of the \al-rich stars, we have revised the stellar parameters of the three RGB stars KIC~3455760, KIC~4350501, and KIC~ 9821622 including the evolutionary stage and composition information. Initially, we repeated the grid-based analysis using the same statistical procedure and set of BaSTI isochrones as in \cite{Pinsonneault2014}, but in this case taking into account \al-enhancement (with the prescription of \citealp{Salaris1993}) and the additional constraint of the period spacing to force the solution to the appropriate evolutionary phase (see \citealp{SilvaAguirre2014}, \citealp{Casagrande2014} for details).
The masses obtained including this new set of information are compatible with those from the original catalogue and thus favour a young age, with the uncertainties in mass being reduced due to the inclusion of the evolutionary phase information.

Nevertheless, recent results suggest that red-giant stellar parameters determined from grid-based analysis might be biased when compared to independent measurements of interferometric radius \citep{Johnson2014}. As a further check on the obtained masses, we have computed models using the GARching STellar Evolution Code (GARSTEC, \citealp{Weiss2008}) to predict the variations of period spacing in the two RGB stars where those measurements are available. The theoretical determinations of $\Delta\Pi_1$ have been made using the asymptotic formulation (see e.g., \citealp{Tassoul1980}).

In the case of KIC~3455760 the period spacing value from Table~\ref{tab:type} is compatible with a star of mass $\sim$1.3~\msun and age $\sim$5~Gyr, with a secondary solution at a higher mass ($\sim$2.0~\msun and age $\sim$1~Gyr). The uncertainty in $\Delta\Pi_1$ prevents a more precise determination of the stellar parameters but it confirms the young nature of the target. For KIC~4350501 the measured period spacing is slightly lower than that predicted by models,
but still favours young stars with masses above $\sim$1.6~\msun, at the high-end of the 1--$\sigma$ uncertainty determined from grid-based modelling.

\begin{table}
\begin{center}
\caption{Asymptotic period spacing and stellar classification from Mosser at al. (2012, 2014) for the young \al-rich stars (in the same order as in Table \ref{tab:ages})}\label{tab:type}
\begin{tabular}{lcc}
\hline
\hline
KIC ID &$\Delta \Pi_1$ [s] & Type\\
\hline
9821622 	 	& 	--- 		& RGB \\
4143460 		& 	287.2 $\pm$ 0.5 & clump\\
4350501 	 	& 	69.3 $\pm$ 0.1	& 	RGB\\
11394905 	 		& 298.0 $\pm$ 0.4 		& clump\\
9269081 	 	& 	351.5 $\pm$ 0.6 	& 	subflash\\
11823838 	  	& 335.7 $\pm$ 0.4 	& 	clump\\
5512910 		& 	333.3 $\pm$ 0.4 & clump\\
10525475 	 	& 	323.9 $\pm$ 0.4 		& clump\\
9002884 	 	& 	--- 		& RGB/AGB\\
9761625 		& 	---		& RGB/AGB\\
11445818 		& 	307 $\pm$ 4 		&  clump\\
3455760 	& 	64.3 $\pm$ 3		& RGB\\
8547669 		& 	 --- 	& 	RGB/AGB\\
3833399 	 	& 	298.5 $\pm$ 3.5 		& clump\\
\hline
\end{tabular}
\end{center}
\end{table}

\begin{figure}
\centering 
\includegraphics[width=0.5 \textwidth]{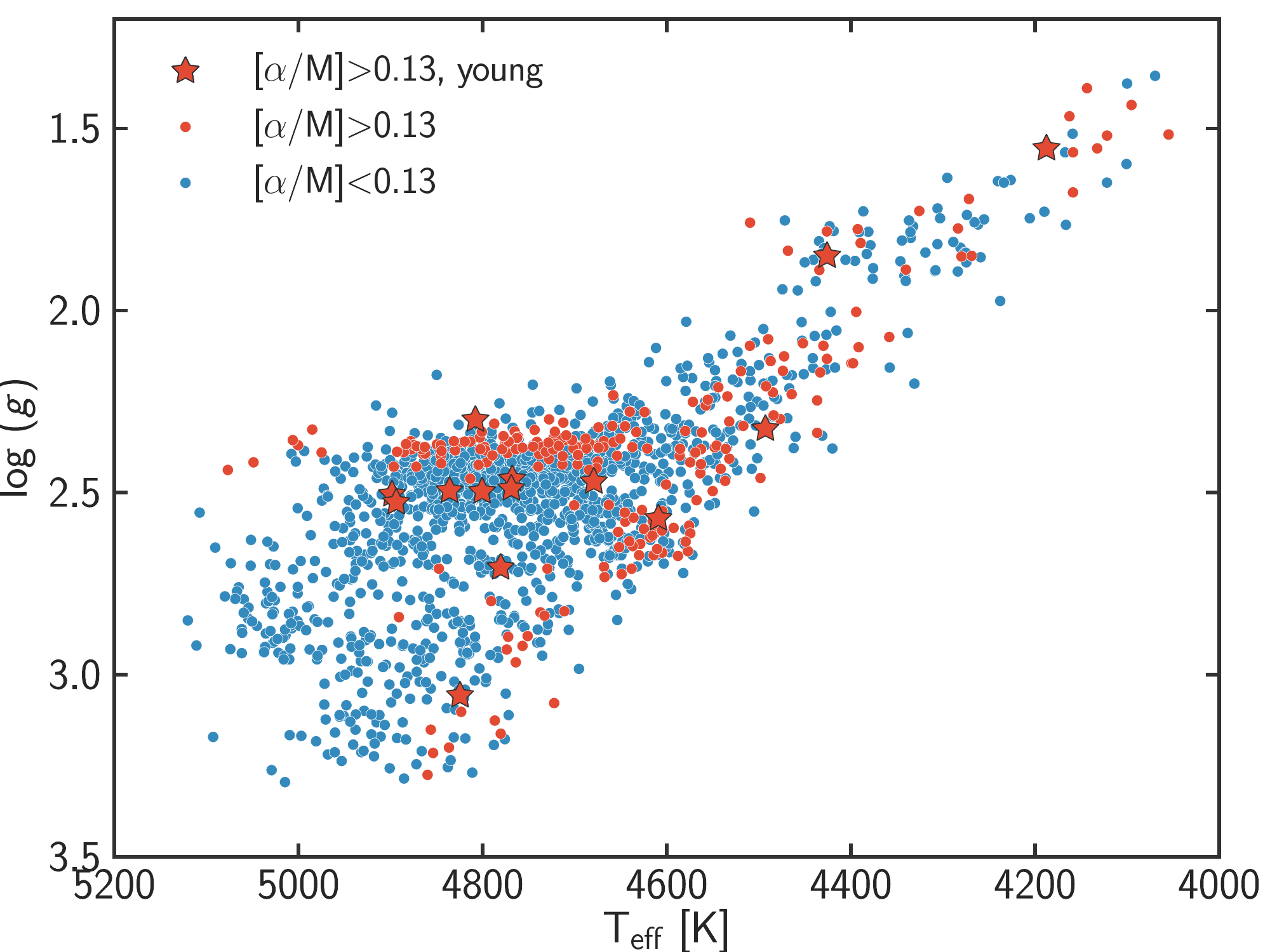}
\caption{Surface gravity as a function of \teff for the APOKASC sample, showing the location of \al-poor stars (blue dots), \al-rich stars (red dots) and \al-rich young stars (red stars). The young \al-rich stars lie at higher log($g$) on the red clump compared to other \al-rich stars, consistently with predictions from theoretical isochrones (see Figure \ref{fig:HR_rich_isochrones}).}
\label{fig:HR_rich}
\end{figure}
\begin{figure}
\centering 
\includegraphics[width=0.5 \textwidth]{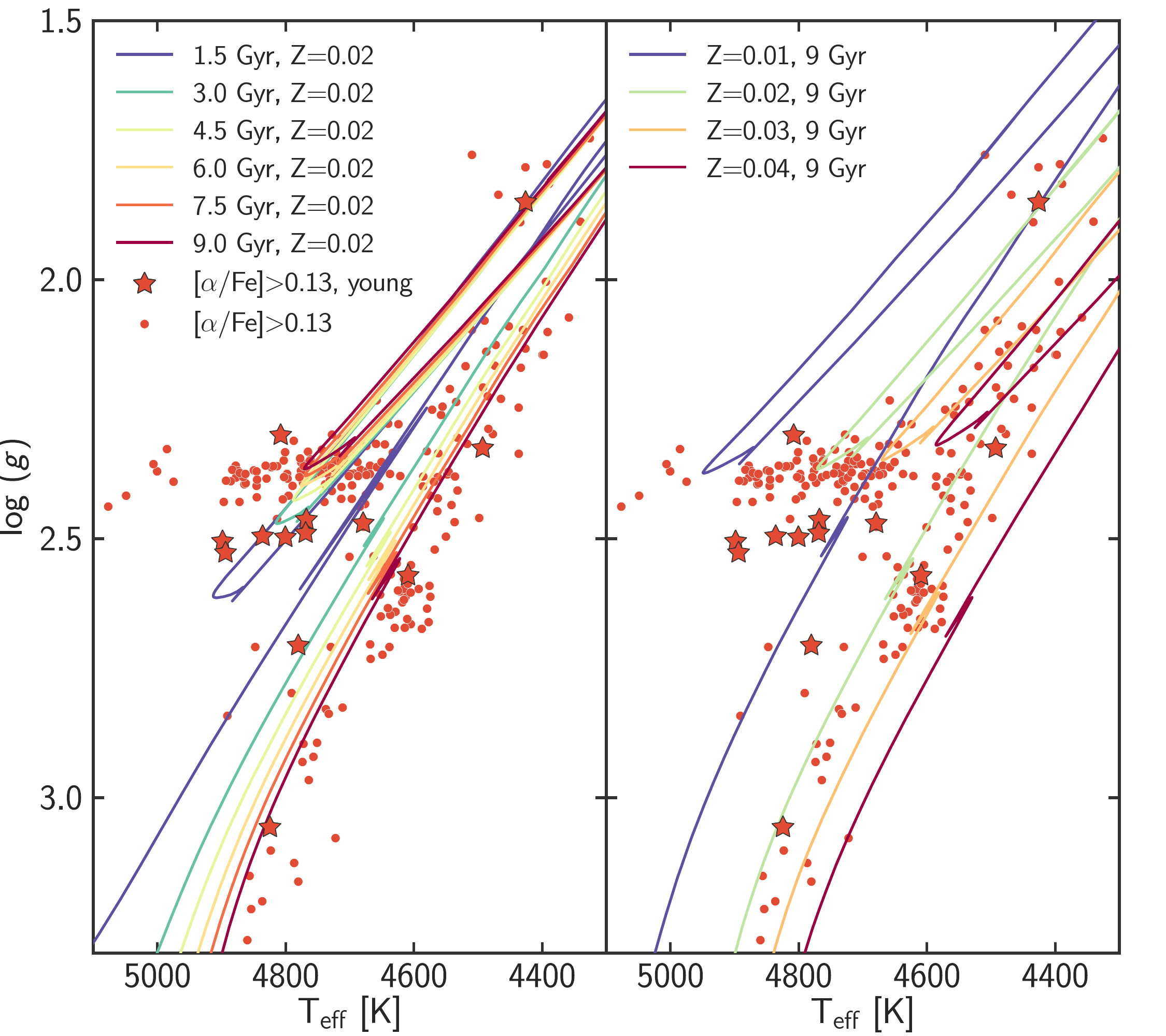}
\caption{Comparison of the location of the \al-rich stars in an H-R diagram with theoretical isochrones. We show BaSTI non-canonical \al-enhanced isochrones for $\eta=0.2$, in the left panel for $Z=0.02$ and ages 1.5, 3, 4.5, 6, 7.5 and 9 Gyr (from blue to red color), and in the right panel for $Z=0.01$, 0.02, 0.03 and 0.04, and an age of 9 Gyr (from blue to red color). These isochrones demonstrate that in the red clump younger stars should have a higher log($g$), which is consistent with our findings. The particular set of isochrones shown on the left panel do not match the location of most \al-rich stars on the RGB: these stars are better matched with isochrones for a slightly higher metallicity (right panel).}
\label{fig:HR_rich_isochrones}
\end{figure}

We thus have additional evidence for correctly inferred high masses for some  of the \al-rich stars in our sample. The next critical step is the translation of mass to age. We have shown that our maximal ages are quite robust versus changes of stellar evolution model (see Figure \ref{fig:stages}), or versus changes in the helium content of the stars (Figure \ref{fig:helium}).

A remaining issue is the question whether a significant fraction of the apparently young, \al-rich stars are massive because they accreted mass from a binary companion or are the result from a stellar merger. In this case, their current mass would not reflect their evolutionary state and their age. Such over-massive red giants in binary systems have for instance been discovered in the open cluster NGC~6819 \citep{Corsaro2012,Brogaard2014}.

Stars that accrete mass from a  companion or that result from a merger appear as blue stragglers when they are on the main sequence. They are easily detected in globular clusters, where they appear bluer than the main sequence turn-off. Evolved blue stragglers are more challenging to identify, they are slightly bluer than normal stars on the RGB, and 0.2 to 1 magnitude brighter on the horizontal branch (HB) \citep{Ferraro1999, Sills2009}. What matters for our study is the fraction of \al-rich stars that could be evolved blue stragglers.

A Hubble Space Telescope study of the populations of blue stragglers, HB and RGB stars in globular clusters by \cite{Leigh2011} shows that the fraction of blue stragglers relative to HB and RGB stars varies from 5 to 15\%. The relative abundance of evolved blue stragglers compared to "normal" blue stragglers is about 1 to 10 \citep{Ferraro1999, Sills2009}. This means that the fraction of evolved blue stragglers to giant stars is of the order of 0.5 to 1.5\%.

For the 241 \al-rich stars in our APOKASC sample, we could then expect 1--4 evolved blue stragglers (using the frequency of blue stragglers observed in globular clusters). Another useful estimate is provided by the \textit{Kepler}  study by \cite{Corsaro2012}   of 115 red giants in the three open clusters, NGC 6791, NGC 6811, and NGC 6819. Amongst these 115 RGB stars (in very similar environments to our \al-rich stars), they find two over-massive stars, that they argue are evolved blue stragglers. If we extrapolate this to our APOKASC sample, we could expect about twice as many, i.e. four in total, evolved blue stragglers.

We have limited the contamination of our sample by such stars as we have removed all stars with anomalous surface rotation as identified by \cite{Tayar2015}. In addition, none of the 14 stars display any peculiar behavior in their \textit{Kepler} light curve; following \cite{Garcia2014} and Ceillier et al. (in preparation), we were unable to determine any signature of surface rotation on these stars from the \textit{Kepler} data up to 100 days period.

\cite{Anders2014} have also used the scatter in radial velocity between successive APOGEE observations of a given star to eliminate binaries (assuming that binaries are found at $\sigma_{v}>1$ \kms). All our \al-rich young stars have a $\sigma_{v}$ below 0.2 \kms from multiple APOGEE observations, which further reduces their probability of being binary stars.
This does not exclude the presence of a few evolved blue stragglers in our sample, but these stars should be too rare to explain the nature of our 14 over-massive stars.

\begin{figure*}
\centering 
\includegraphics[width=\textwidth]{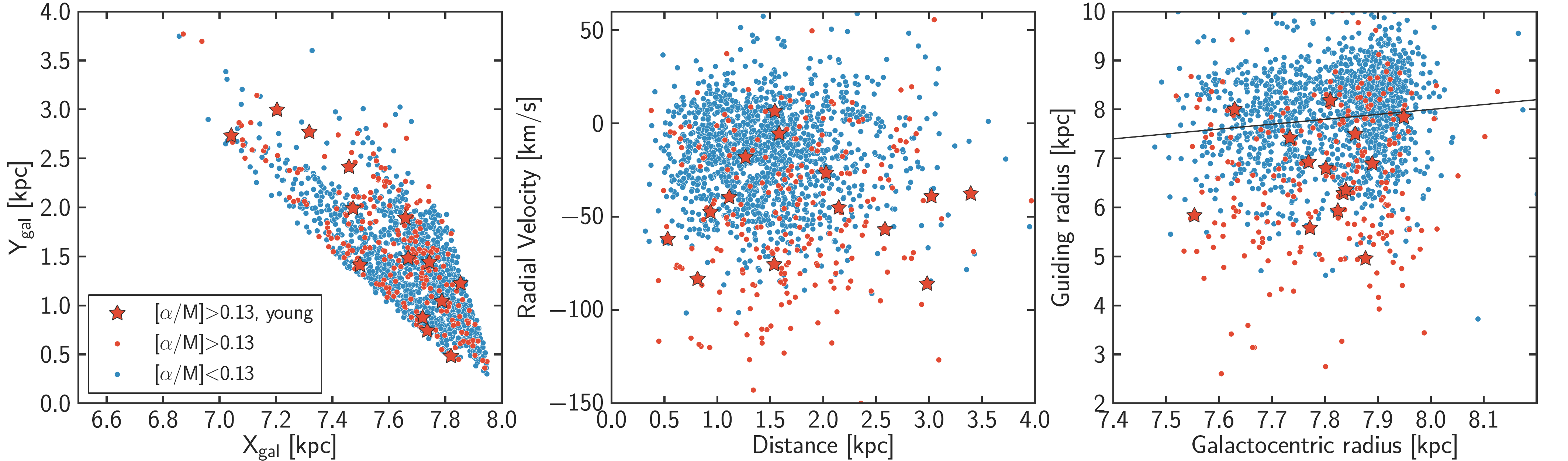}
\caption{Spatial distribution, radial velocity, and guiding radius for  \al-poor stars (blue dots), \al-rich stars (red dots) and \al-rich young stars (red stars). The distances to the Sun have been computed by Rodrigues et al. (2014), and are used to derive galactocentric coordinates $X_{\mathrm{gal}}$ and $Y_{\mathrm{gal}}$, and galactocentric radius $R_{\mathrm{gal}}$. The guiding radii is computed as $R_{\mathrm{guide}}=L_z / V_{\mathrm{circ}}=R_{\mathrm{gal}}V_{\phi} /  V_{\mathrm{circ}}$, using $V_{\mathrm{circ}}=220$ \kms and proper motions from the UCAC4 catalogue \citep{Zacharias2013}. The \al-rich young stars have orbital properties similar to the rest of the \al-rich population.}
\label{fig:spatial_dist}
\end{figure*} 
Finally, another line of evidence supporting the young nature of some of our stars is their location in the H--R diagram (see Figure \ref{fig:HR_rich}). Because of \teff uncertainties, it is difficult to assess the age of stars on the RGB or AGB through comparisons with isochrones. For clump stars, however, there is a dependence of log($g$) on age. At fixed metallicity, young (massive)  clump stars have a higher log ($g$) than older stars (as seen from the \al-enhanced BaSTI isochrones plotted in Figure \ref{fig:HR_rich_isochrones}). Amongst \al-rich stars, those that we identified as young indeed have a significantly higher log ($g$) compared to the ''normal'' old \al-rich stars, with an offset roughly consistent with ages below 3 Gyr. In this Figure, the \al-rich young star that appears to be a clump star but has a low log ($g$) of 2.3 is actually not yet in the clump: it is the star whose period spacing suggests it might be in the subflash phase.

\section{Discussion and conclusion}

A careful analysis, exploring many potential modelling systematics, has brought us to infer that  the APOKASC sample contains at least 14 stars that are both enriched in \al-elements and younger than 6 Gyr. Our approach is as conservative as possible, so that there might be more young \al-rich stars in the sample, which we have not identified. The ages could be slightly higher if the standard seismic scaling relations would have to be revised, or if the stars's helium content is very low. None of these options can make the 14 stars older than 8--9 Gyr, which is the age commonly found for \al-rich stars in the Milky Way \citep[see for instance the recent results by][]{Haywood2013,Bensby2014,Bergemann2014}.

Such young \al-rich stars are similarly found by Anders et al. (in preparation) and \cite{Chiappini2015} in the CoRoT-APOGEE (CoRoGEE) sample, with ages determined from grid-based modelling using asteroseismic information.  As described in more detail in \cite{Chiappini2015}, a few young \al-rich stars are also actually present in the samples described in \cite{Haywood2013}, \cite{Bensby2014} and \cite{Bergemann2014}, using very different age determination techniques, and \al-element abundances from high-resolution optical spectra. 

Altogether, the combination of evidence from different studies confirms the existence of this population of  stars. In CoRoGEE, \cite{Chiappini2015}  find comparatively more  \al-rich young stars at small galactocentric radii compared to CoRoT fields in the outer disc. This result  cannot be tested with APOKASC, where all stars are at a nearly constant galactocentric radius (see Figure \ref{fig:spatial_dist}). Nonetheless, any radial trend would be difficult to explain if such stars were just an artefact in the data.

This combination of high \aFe and young ages is not predicted by standard chemical evolution models of the Galaxy.  In the models presented in \cite{Minchev2013} (their Figure 2), stars with [O/Fe]=0.2 are older than 7 Gyr, whatever their birth location. Stars with slightly smaller [O/Fe] can be younger at the condition that they are born in the very outer disk, but they would then have metallicities below $-0.5$, which is lower than what we find. \cite{Chiappini2015} suggest that such stars could be born at the end of the bar and then  migrate to the solar radius, but it is still unclear if this scenario is  a valid one. One expectation in such a scenario would be that the \al-rich young stars have smaller guiding radii than the rest of the population.
Figure \ref{fig:spatial_dist} displays the radial velocity and guiding radius distribution for the APOKASC sample. The \al-rich and \al-poor stars have clearly different orbital properties, with an average guiding radius of 6.5 kpc for the \al-rich stars versus 7.9 kpc for the \al-poor stars, and average radial velocities of $-50$ \kms versus $-17$ \kms, respectively. However, the \al-rich young stars do not possess distinct orbital properties compared to the rest of the \al-rich population, with an average guiding radius of 6.7 kpc and an average radial velocity of $-44$ \kms. Thus, we cannot confirm a different birth location for the \al-rich young stars compared to the other \al-rich stars.

More detailed determination of their orbital properties, and comparisons with new galactic chemical evolution models, as well as combined studies with the CoRoGEE sample, might  shed light on their possible origin. What appears clear is that, if they are truly young, they cannot have formed at the solar radius, and they thus form a  sample of stars having experienced radial migration, although from an unknown location. While \al-enrichment correlates quite well with age for the general population, \aFe cannot be used blindly as a proxy for age on a star-by-star basis.

\section*{Acknowledgments}
We thank Melissa Ness and Anna Sippel for useful discussions. We thank the referee for a constructive report.
MM acknowledges support from the Alexander von Humboldt Foundation. The research has received funding from the European Research Council under the European Union's Seventh Framework Programme (FP 7) ERC Grant Agreement n. [321035]. Funding for the Stellar Astrophysics Centre is provided by The Danish National Research Foundation (Grant agreement no.: DNRF106). The research leading to the presented results has received funding from the European Research Council under the European Community's Seventh Framework Programme (FP7/2007-2013)/ERC grant agreement no 338251 (StellarAges). The research is supported by the ASTERISK project (ASTERoseismic Investigations with SONG and Kepler) funded by the European Research Council (Grant agreement no.: 267864). BM acknowledges financial support from the Programme National de Physique Stellaire (CNRS/INSU) and with RAG acknowledge the ANR (Agence Nationale de la Recherche, France) program IDEE (ANR-12-BS05-0008) "Interaction Des \'Etoiles et des Exoplan\'etes". YE acknowledges support from the UK Science and Technology Research Council. DS acknowledges support from the Australian Research Council. TSR acknowledges support from CNPq-Brazil.  RC acknowledge financial support provided by the Spanish Ministry of Economy and Competitiveness under grant AYA2010-16717.
MP and JAJ acknowledge support from NSF grant AST-1211673. 
AS is supported by the MICINN grants AYA2011-24704 and ESP2013-41268-R. SM acknowledges the support of the NASA grant NNX12AE17G.
The research leading to these results has received funding from the
the UK Science and Technology Facilities Council (STFC). 
D.A.G.H. and O.Z. acknowledge support provided by the Spanish Ministry of
Economy and Competitiveness under grant AYA-2011-27754.
SB acknowledges partial support from NSF grant AST-1105930 and NASA grant NNX13AE70G. This work has made use of BaSTI web tools.

Funding for SDSS-III has been provided by the Alfred P. Sloan Foundation, the Participating Institutions, the National Science Foundation, and the U.S. Department of Energy Office of Science. The SDSS-III Web site is http://www.sdss3.org/. SDSS-III is managed by the Astrophysical Research Consortium for the Participating Institutions of the SDSS-III Collaboration including the University of Arizona, the Brazilian Participation Group, Brookhaven National Laboratory, Carnegie Mellon University, University of Florida, the French Participation Group, the German Participation Group, Harvard University, the Instituto de Astrofisica de Canarias, the Michigan State/Notre Dame/JINA Participation Group, Johns Hopkins University, Lawrence Berkeley National Laboratory, Max Planck Institute for Astrophysics, Max Planck Institute for Extraterrestrial Physics, New Mexico State University, New York University, Ohio State University, Pennsylvania State University, University of Portsmouth, Princeton University, the Spanish Participation Group, University of Tokyo, University of Utah, Vanderbilt University, University of Virginia, University of Washington, and Yale University.

{}

\end{document}